\documentclass[12pt,preprint]{aastex}

\newcommand{\etal}{et~al.\/}

\newcommand{\dms}[3]{\ensuremath{{#1}\ {#2}\ {#3}}}
\newcommand{\hms}[3]{\ensuremath{{#1}\ {#2}\ {#3}}}

\newcommand{\sqdeg}{square degree}
\newcommand{\sqdegs}{square degrees}
\newcommand{\bootes}{Bo\"otes}

\newcommand{\mjypbm}{{\rm mJy~beam}\ensuremath{^{-1}}}

\newcommand{\ie}{i.\,e.\,}
\newcommand{\eg}{e.\,g.\,}

\newcommand{\uv}{\ensuremath{(u,v)}}
\newcommand{\satats}{\ensuremath{S_{\rm ATATS}}}
\newcommand{\snvss}{\ensuremath{S_{\rm NVSS}}}
\newcommand{\smi}{\ensuremath{S_{m_i}}}
\newcommand{\sci}{\ensuremath{S_{c_i}}}
\newcommand{\sma}{\ensuremath{\bar{S}_{m}}}
\newcommand{\sca}{\ensuremath{\bar{S}_{c}}}
\newcommand{\postsize}{Postage stamps are $30\arcmin \times 30\arcmin$, centered on the ATATS position. The 150\arcsec\ circular Gaussian restoring beam is shown in the lower right of the ATATS image. The integrated flux density of the ATATS source at the center of the image, and the integrated flux density of the closest NVSS source brighter than 10\,mJy within 75\arcsec, both in mJy, are shown above the ATATS and NVSS images respectively. The greyscale runs from $-20$ to $+50$\,\mjypbm.}
\newcommand{\postsizetwo}{Postage stamps are $30\arcmin \times 30\arcmin$, centered on the ATATS position. The 150\arcsec\ circular Gaussian restoring beam is shown in the lower right of the ATATS image. The integrated flux density of the ATATS source at the center of the image is shown above the ATATS image. Here there is no NVSS source brighter than 10\,mJy within 75\arcsec, so the NVSS flux density is labelled ``$f=0$''. The greyscale runs from $-20$ to $+50$\,\mjypbm.}

\shorttitle{The ATA 20-cm Survey}
\shortauthors{Croft \etal}

\begin{document}
\title{The Allen Telescope Array Twenty-centimeter Survey -- A 690-Square-Degree, 12-Epoch Radio Dataset -- I: Catalog and Long-Duration Transient Statistics }
\author{Steve Croft\altaffilmark{1}, Geoffrey C.\ Bower\altaffilmark{1},
 Rob Ackermann\altaffilmark{2},
 Shannon Atkinson\altaffilmark{2},
 Don Backer\altaffilmark{1},
 Peter Backus\altaffilmark{2},
 William C. Barott\altaffilmark{3},
 Amber Bauermeister\altaffilmark{1},
 Leo Blitz\altaffilmark{1},
 Douglas Bock\altaffilmark{1},
 Tucker Bradford\altaffilmark{2},
 Calvin Cheng\altaffilmark{1},
 Chris Cork\altaffilmark{4},
 Mike Davis\altaffilmark{2},
 Dave DeBoer\altaffilmark{5},
 Matt Dexter\altaffilmark{1},
 John Dreher\altaffilmark{2},
 Greg Engargiola\altaffilmark{1},
 Ed Fields\altaffilmark{1},
 Matt Fleming\altaffilmark{4},
 James R.~Forster\altaffilmark{1},
 Colby Gutierrez-Kraybill\altaffilmark{1},
 Gerry Harp\altaffilmark{2},
 Tamara Helfer\altaffilmark{1},
 Chat Hull\altaffilmark{1},
 Jane Jordan\altaffilmark{2},
 Susanne Jorgensen\altaffilmark{1},
 Garrett Keating\altaffilmark{1},
 Tom Kilsdonk\altaffilmark{2},
 Casey Law\altaffilmark{1},
 Joeri van Leeuwen\altaffilmark{6},
 John Lugten\altaffilmark{7},
 Dave MacMahon\altaffilmark{1},
 Peter McMahon\altaffilmark{8},
 Oren Milgrome\altaffilmark{1},
 Tom Pierson\altaffilmark{2},
 Karen Randall\altaffilmark{2},
 John Ross\altaffilmark{2},
 Seth Shostak\altaffilmark{2},
 Andrew Siemion\altaffilmark{1},
 Ken Smolek\altaffilmark{2},
 Jill Tarter\altaffilmark{2},
 Douglas Thornton\altaffilmark{1},
 Lynn Urry\altaffilmark{1},
 Artyom Vitouchkine\altaffilmark{4},
 Niklas Wadefalk\altaffilmark{9},
 Jack Welch\altaffilmark{1},
 Dan Werthimer\altaffilmark{1},
 David Whysong\altaffilmark{1},
 Peter K.~G.~Williams\altaffilmark{1}, and
 Melvyn Wright\altaffilmark{1}}
\altaffiltext{1}{University of California, Berkeley, 601 Campbell Hall \#3411, Berkeley, CA 94720, USA }
\altaffiltext{2}{SETI Institute, Mountain View, CA 94043, USA }
\altaffiltext{3}{Embry-Riddle Aeronautical University, Electrical, Computer, Software, and Systems Engineering Department, Daytona Beach, FL 32114, USA }
\altaffiltext{4}{Minex Engineering, Antioch, CA 94509, USA }
\altaffiltext{5}{CSIRO/ATNF, Epping, NSW 1710, Australia }
\altaffiltext{6}{ASTRON, 7990 AA Dwingeloo, The Netherlands}
\altaffiltext{7}{Lawrence Livermore National Laboratory, Livermore, CA 94550, USA}
\altaffiltext{8}{Electrical Engineering Department, Stanford University, Stanford, CA 94305, USA }
\altaffiltext{9}{Chalmers University of Technology, Department of Microtechnology and Nanoscience - MC2, SE-412 96 G\"{o}teborg, Sweden} 

\tabletypesize{\scriptsize}

\begin{abstract}

We present the Allen Telescope Array Twenty-centimeter Survey (ATATS), a multi-epoch (12 visits), 690~\sqdeg\ radio image and catalog at 1.4\,GHz. The survey is designed to detect rare, very bright transients as well as to verify the capabilities of the ATA to form large mosaics. The combined image using data from all 12 ATATS epochs has RMS noise $\sigma = 3.94$\,\mjypbm\ and dynamic range 180, with a circular beam of 150\arcsec\ FWHM. It contains 4408 sources to a limiting sensitivity of $5\sigma = 20$\,\mjypbm. We compare the catalog generated from this 12-epoch combined image to the NRAO VLA Sky Survey (NVSS), a legacy survey at the same frequency, and find that we can measure source positions to better than $\sim 20$\arcsec. For sources above the ATATS completeness limit, the median flux density is 97\%\ of the median value for matched NVSS sources, indicative of an accurate overall flux calibration. We examine the effects of source confusion due to the effects of differing resolution between ATATS and NVSS on our ability to compare flux densities. We detect no transients at flux densities greater than 40\,mJy in comparison with NVSS, and place a $2 \sigma$ upper limit on the transient rate for such sources of 0.004\,deg$^{-2}$. These results suggest that the $\gtrsim 1$\,Jy transients reported by \citet{matsumura:09} may not be true transients, but rather variable sources at their flux density threshold.

\end{abstract}


\section{Introduction}

The Allen Telescope Array (ATA), a joint project of the Radio Astronomy Laboratory of the University of California, Berkeley, and the SETI Institute in Mountain View, CA, is a radio telescope built on a novel ``LNSD'' (Large Number of Small Dishes) design \citep{welch}.  It is designed to have a wide field of view ($\sim$5~\sqdegs\ at 1.4\,GHz), enabling rapid surveys of large areas of sky. This makes it an ideal instrument to search for transient and time-varying radio sources. Since survey speed (the area of sky covered to a given flux density limit in a fixed amount of time) is proportional to $ND$, the product of the number of dishes, $N$, and their diameter, $D$ \citep{sargent}, this gives the ATA a survey speed $\sim 40$\%\ that of the Very Large Array. Plans call for the ATA's expansion from the existing 42 6\,m-diameter antennas to a total of 350 antennas, which will result in survey speed an order of magnitude faster still than the current configuration. 

The technical design (including the novel antennas, feeds, and correlator), and some of the early science results, are described by \citet{welch}. Additional early science is described by \citet{williams} as well as other papers in preparation. In this paper we present the ATA Twenty-centimeter Survey (ATATS), which consists of observations of 690~\sqdegs\ of sky centered on the $\sim 10$~\sqdeg\ \bootes\ deep field \citep{ndwfs}. This is a region with a wealth of multi-wavelength data, including data on mid-infrared variability \citep{sdwfsvar}, which we will compare with radio variability from ATATS in a future paper.

ATATS observations were spread over 12 epochs from 2009 January -- April (Table~\ref{tab:epochs}). Each epoch we aimed to observe as many of the 361 ATATS pointings (\S~\ref{sec:design}) as possible; in practice, at a given epoch, good data were obtained for between 252 and 318 pointings (Table~\ref{tab:epochs}). Observing efficiency and telescope uptime have improved since the ATATS data were taken. Occasionally a given pointing failed for some reason -- for example, strong radio frequency interference (RFI) -- and so not all epochs have good data for all pointings observed. A few observation attempts resulted in corrupted data or no data at all, due to software, hardware, or calibration problems. Epochs with a significant number of failed pointings were excluded from the dataset presented in this paper.

\begin{deluxetable}{lllll}
\tablewidth{0pt}
\tabletypesize{\scriptsize}
\tablecaption{\label{tab:epochs} ATATS epochs}
\tablehead {
\colhead{Epoch} & \colhead{UT Date} & \colhead{Pointings} 
}
\startdata
ATA1  & 2009 Jan 12 & 255 \\
ATA2  & 2009 Jan 19 & 271 \\
ATA3  & 2009 Jan 26 & 254 \\
ATA4  & 2009 Jan 31 & 285 \\
ATA5  & 2009 Feb 7  & 285 \\
ATA6  & 2009 Feb 14 & 318 \\
ATA7  & 2009 Feb 15 & 310 \\
ATA8  & 2009 Mar 27 & 261 \\
ATA9  & 2009 Mar 28 & 252 \\
ATA10  & 2009 Mar 29 & 270 \\
ATA11 & 2009 Mar 31 & 257 \\
ATA12 & 2009 Apr 3  & 253 \\
\enddata
\end{deluxetable}

This paper examines the image and catalog made with data from all 12 epochs, and compares it to the NRAO VLA Sky Survey \citep[NVSS;][]{nvss}. The analysis of the images of individual epochs, including transient statistics obtained from comparing catalogs made from data taken at a single epoch, will be the subject of \citet{paperii}.

Throughout this paper, we use J2000 coordinates.

\section{Transients and Variables at Radio Wavelengths}

ATATS was designed to verify the performance of the ATA during commissioning (by comparison to existing sky survey data) as well as to search for time-varying and transient sources. The survey was designed to cover large areas at a comparatively low sensitivity. 

Here we distinguish between sources which are normally above the flux density limit (which we term ``variable'' and ``non-variable'', depending on whether or not changes in their flux density are detectable from epoch to epoch -- although most radio sources are intrinsically variable to some degree) and those which have no quiescent radio counterpart (which we term ``transient''). There is obviously some overlap between transient and variable sources, in that some sources which exhibit large outbursts can be seen in quiescent radio emission if observed to faint enough flux densities. But it is useful (particularly when comparing catalogs taken at different epochs) to distinguish between variable sources which are present in most or all epochs (often rotating sources, or sources with variable accretion rates or magnetic fields), and transients which appear above the detection limit at one or a few epochs, and are undetected at others (often explosive events). Variable and transient sources manifest on a range of different timescales in the radio.

{\em Transients} have been observed with timescales from milliseconds \citep{lorimer:07} to months \citep{m82sn}, from a range of progenitors. These include gamma ray bursts \citep{frail:00} and radio supernovae \citep{m82sn} -- where the progenitor is destroyed -- and flares from a variety of objects, which may or may not also show up as variable sources during their more quiescent periods. Flares are seen from active galactic nuclei \citep[AGNs;][]{falcke:99} at one end of the size scale, to M-dwarfs \citep{jackson:89}, brown dwarfs \citep{berger:01,hallinan:07}, pulsars \citep{cognard:96}, and rotating radio transients \citep[RRATs;][]{rrats} at the other.

Some radio transients \citep{bower,matsumura:09}\label{sec:archival} show no counterparts in very deep observations at other wavelengths; these may be flares from isolated old neutron stars \citep{ofek:10} or perhaps something more exotic. A few may not even be astrophysical in origin at all -- the \citet{lorimer:07} transients are now thought to perhaps be due to terrestrial lightning \citep{bs}, although astrophysical interpretations are not yet completely ruled out.

{\em Variable sources} at radio wavelengths are also associated with several different classes of object and characteristic timescale. Stars often exhibit variable radio emission \citep{gudel:02}, as do pulsars \citep{archibald}, microquasars \citep{mj:09}, and AGNs.

AGNs make up almost all of the sources seen in imaging surveys above a few mJy at 1.4\,GHz \citep{seymour:08}, and vary over a range of timescales. Some AGNs exhibit IntraDay Variability (IDV) -- scintillation of extremely compact components due to extrinsic sources, likely located in our own Galaxy \citep{lazio,microlensing}. This can cause flux changes ranging from a few percent, up to 50\%\ in so-called Extreme Scattering Events \citep{ese}.  

On timescales of months, radio synchrotron emission is sometimes seen to vary in correlation with optical \citep{arshakian:10} or X-ray \citep{marscher:02} emission, as uneven accretion rates from the in-spiraling disk propagate through to knots in the outflowing jets; on timescales of years, even more substantial variations in accretion rate, heating of material, and reprocessing of energy released in the accretion disk can occur. 

Long-term variability of very bright AGNs ($S \gtrsim 1$\,Jy) has been systematically explored \citep[\eg,][]{aller:92},  but there has not been systematic characterization of the fainter AGN population.  Some faint AGNs such as III~Zw~2 show extreme variations \citep{falcke:99}.
And over periods of $\sim 10^7$\,yr, quasar activity is thought to switch on and off \citep{hopkins}, although the duty cycle is not well understood and accretion may stop and start several times over such periods \citep{saikia}.

The cadence, frequency, and sensitivity of the observations influences what kind of variable and transient objects will be seen in a given survey. We selected an extragalactic field, which we observed at 1.4\,GHz to relatively low sensitivity, with a cadence of a few days to weeks, so almost all variable and non-variable sources seen are AGNs, and any transients would be likely to be radio supernovae or gamma ray bursts, although some sources may be members of other classes of object discussed above, or even more exotic objects.

\section{Survey Design and Data Acquisition} \label{sec:design}

The ATA has 42 dual-polarization antennas, but the ATATS survey was undertaken during a period when the array was still being commissioned, and at times, various feeds were disconnected, being retrofitted (to improve electrical connections in the feed tip circuit board), or otherwise not in operation, so not all antennas, baselines, or polarizations were present in all datasets. In any case, at the time of these observations, the correlator was only capable of calculating correlations of 64 inputs (either 32 dual-polarization feeds, or some other combination of 64 single-polarization inputs from the 42 antennas). Out of a possible 861 baselines (each with two polarizations) the correlator was therefore capable of producing correlations for 496. Due to the fact that not all 64 inputs were populated with good feeds at the time of the observations, the mean number of baselines present in the ATATS dataset is 340, representing an efficiency of 39\%. Since the ATATS data were taken, improvements in the feeds, and the installation of a second 64-input correlator, have increased the efficiency substantially.

Some of the ATATS data were later flagged due to radio frequency interference (RFI) or other problems (\S~\ref{sec:rfi}), reducing the observing efficiency still further. Nevertheless, we were still able to achieve good throughput in terms of the product of the area surveyed and the sensitivity reached, and as noted, the observing efficiency of the ATA continues to improve, and has the potential to improve even more dramatically if more antennas are added to the array.

\begin{figure}
\centering
\includegraphics[width=\linewidth,draft=false]{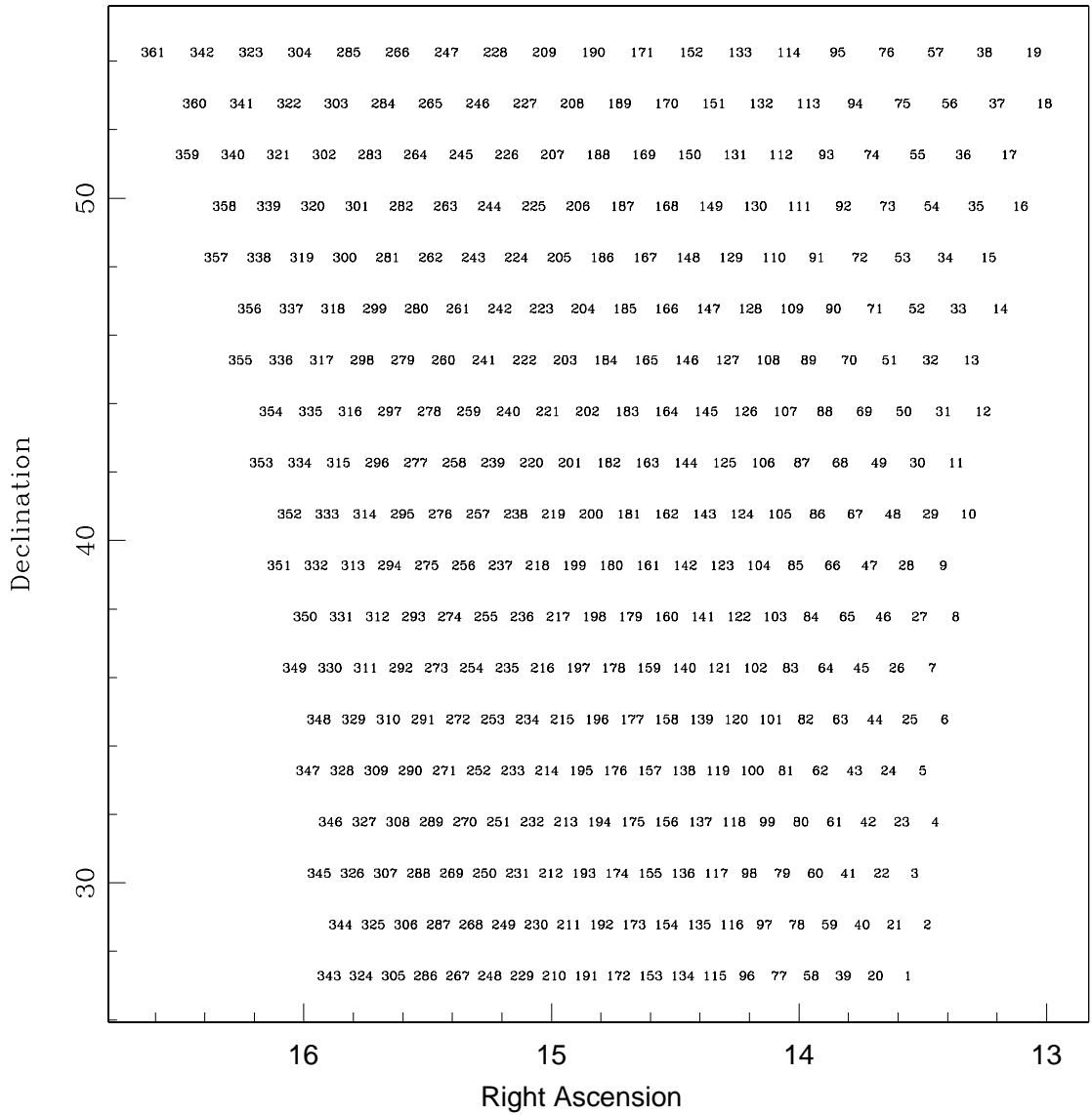}
\caption{\label{fig:pointings}
The field centers of the 361 ATATS pointings.
}
\end{figure}

The layout of the ATATS pointings, and the sequence in which they were observed, is shown in Fig.~\ref{fig:pointings}. Pointings were arranged to approximate a hexagonal close-packed grid, with the spacing between pointing centers determined by Nyquist sampling, and accounting for the convergence of lines of constant right ascension towards the poles. 

Each pointing was imaged once per epoch as a snapshot observation, with an integration time of 60\,s ($6 \times 10$\,s scans). Every 50 minutes, the calibrator 3C\,286 was observed for 60\,s, as well as once at the completion of each epoch. Observing always started with pointing number 1, and continued until the end of the scheduled observing time, usually around 12 hours.

The correlator provided 1024 frequency channels, each of width 102\,kHz, across a 104.9\,MHz-wide band centered at 1430\,MHz. The 100 channels at each edge of the band were flagged a priori because of poor sensitivity due to filter roll-off, so the effective bandwidth was 84.4\,MHz.

\section{Data Reduction}

As the number of antennas and the volumes of data from radio telescopes increase, calibration, imaging, and flagging RFI by hand start to become impractical. As a result, automated routines are required to get the best results from the data. Data reduction was undertaken using a software suite developed for the reduction of ATA data, RAPID \citep[Rapid Automated Processing and Imaging of Data;][]{keating:aas}. RAPID consists of C shell scripts that call MIRIAD \citep{miriad} tasks. It performs identification and excision of RFI, calibrates the visibilities using a primary flux calibrator, and images the individual pointings. It then performs phase self-calibration and amplitude flagging. RAPID currently
operates only on ATA data but is freely available by contacting the authors. Mosaicking, catalog creation, and matching to NVSS are performed by additional custom-written scripts.

\subsection{Flagging, Calibration, and Imaging}

RAPID is able to detect and flag RFI on a channel-by-channel basis by looking for channels that often rise above the mean amplitude for all cross-correlations. RAPID also identifies any problem antennas that show signs of corrupted spectra, and flags them. It then makes an image of the calibrator, using full multifrequency synthesis (to avoid bandwidth smearing), and iteratively tests for phase and amplitude closure, flags data outside of a certain amplitude range, and makes a new image.

Once all baselines fall within phase RMS and closure limits, iteration stops and an image of the calibrator is produced. Baselines that were determined to be bad in the calibrator data are also flagged in the source data. RAPID then makes images of the source fields, and iteratively applies phase self-calibration and amplitude flagging while attempting to maximize the image dynamic range. Once a maximum in dynamic range is achieved, iteration stops - for our images, this results in a median number of CLEAN components per image of 5710, a median RMS of 3.56\,\mjypbm\ (2.2 times the theoretical RMS of 1.6\,\mjypbm), and a median dynamic range of 120. Images were made using natural weighting which gives about a factor 3 lower noise than uniform weighting for the ATA. Superuniform weighting was also tried, but the results were unsatisfactory, both in terms of image quality, and much higher noise than natural weighting.

The data were calibrated using 3C\,286. The calibrator flux density was determined to be 14.679\,Jy, using the \citet{baars} coefficients. We did not perform a polarization calibration. The ATA produces cross-polarization data, but at the time of this work, the calibration and analysis of the cross-polarization data have not been fully tested. 3C\,286 is $\sim 10\%$ polarized, but since both X and Y polarizations are present in the data, the effect of source polarization on Stokes I flux densities is largely cancelled; ignoring polarization calibration will lead to an additional $\sim 1\%$ uncertainty in flux density. This is negligible compared to other sources of error.

The flagged visibility data, the bandpasses, and the final images were all inspected by eye, but usually RAPID does a sufficiently good job (given optimal input parameters) that little human interaction is required. Where artifacts were visible in the final images, or RFI or other bad data remained in the visibilities, additional manual flagging was performed.

\label{sec:rfi}For the 12 good datasets an average of 40.6\%\ of visibilities were flagged -- this includes RFI, the 200 edge channels, and other bad data identified during data reduction. As discussed in \S~\ref{sec:design}, the ATA was operating at only 39\%\ of its maximum capacity during ATATS, so the fraction of data obtained compared to the technical specifications was 23\%. This fraction is now increasing as additional retrofitted feeds are brought online and other technical improvements are made, which will improve the fidelity of future surveys such as PiGSS (the Pi GHz Sky Survey; Bower \etal\ in prep.).

\begin{figure}
\centering
\includegraphics[angle=270,width=0.45\linewidth,draft=false]{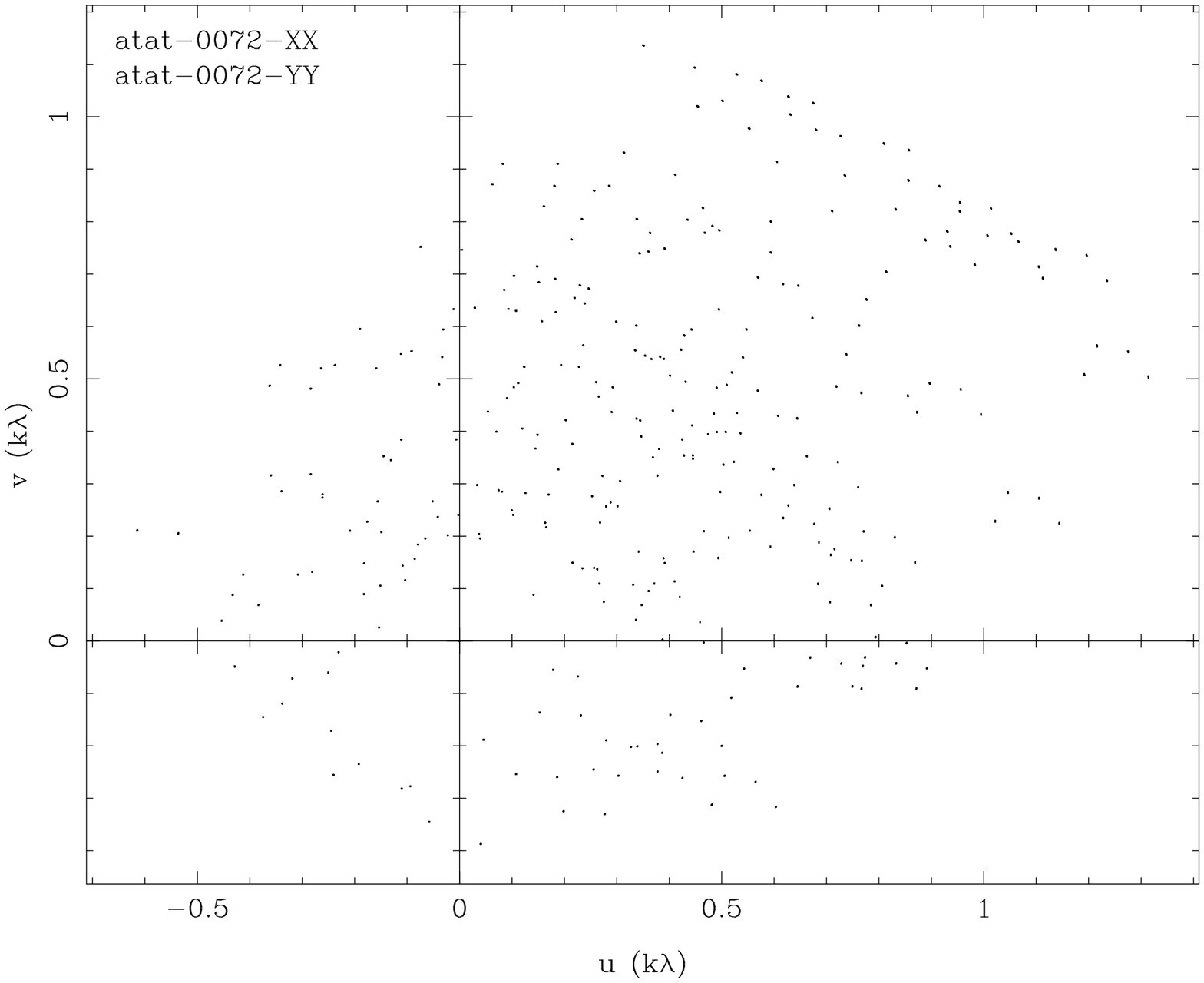}\hspace{0.1\linewidth}%
\includegraphics[angle=270,width=0.45\linewidth,draft=false]{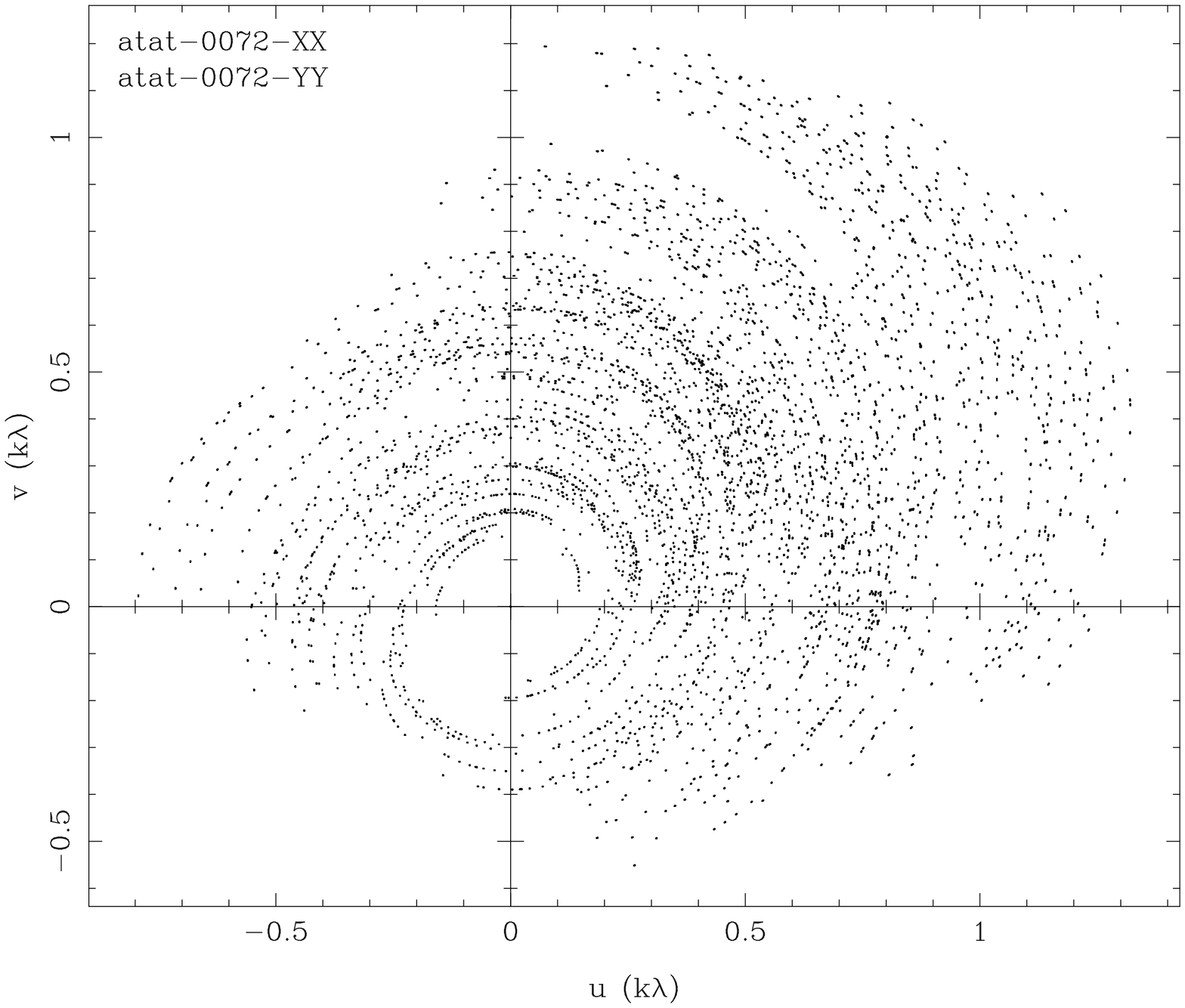}
\caption{\label{fig:uvcover}
Typical \uv\ coverage for a single pointing at a single epoch (left), and for the same pointing when combining the data from all epochs (right).
}
\end{figure}

\begin{figure}
\centering
\includegraphics[angle=270,width=\linewidth,draft=false]{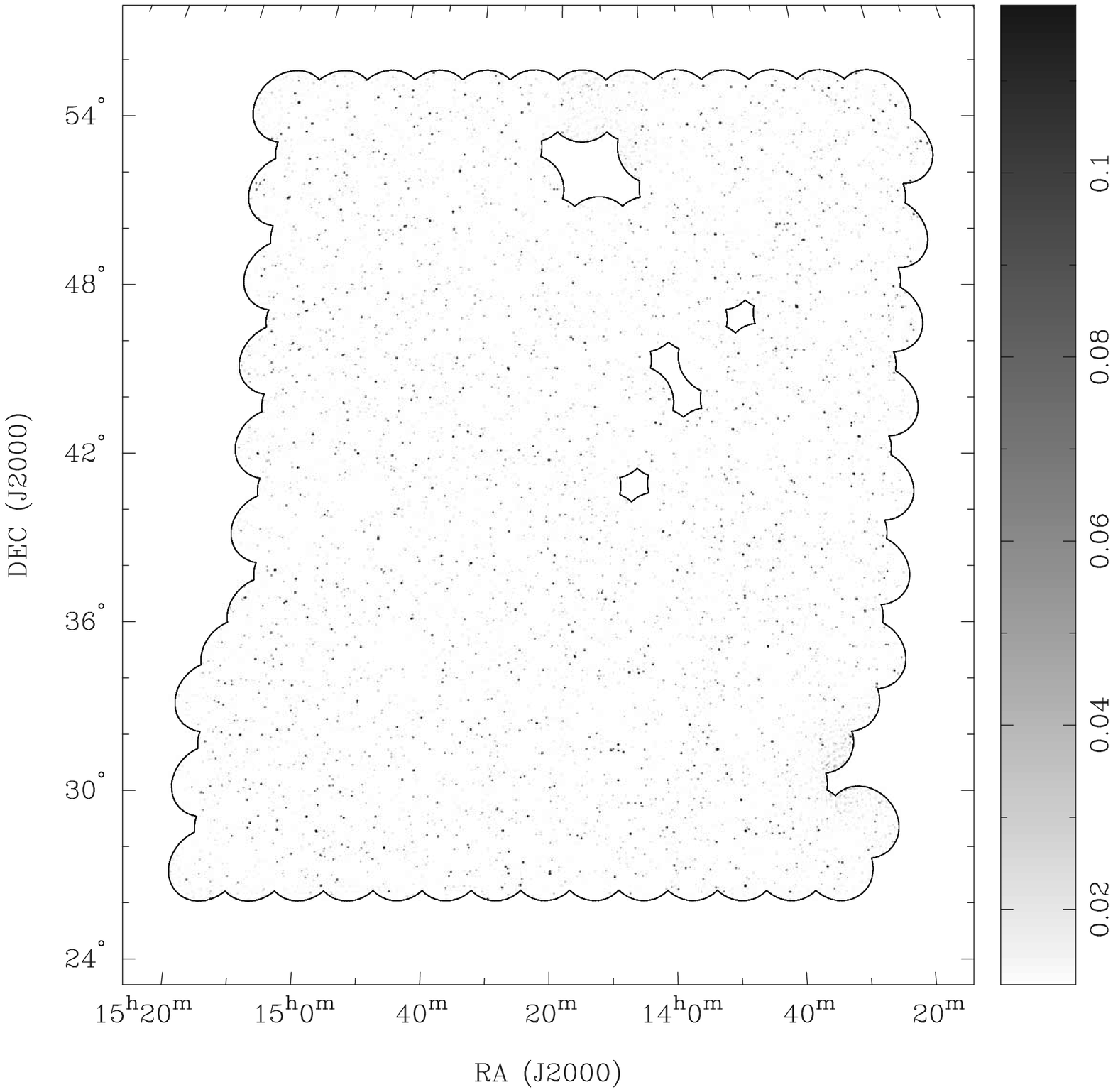}
\caption{\label{fig:deepfield}
The deep field image, made from a combination of all 12 epochs. The solid line shows the edge of the mosaic -- note that some small regions with fewer than 9 epochs of good data were not included in the mosaic. The areas near 3C\,295 (the large missing region at top center) and 3C\,286 (the notch cut into the lower right edge) are notable examples, but there are also a few smaller regions missing, where $> 4$ epochs were unusable due to RFI or some other problem (see \S~\ref{sec:corruptreg}). The greyscale runs from 11.82\,\mjypbm\ ($3 \sigma$) to 118.2\,\mjypbm ($30 \sigma$).
}
\end{figure}

The size of the images made for each pointing was $512 \times 512$ pixels, where each pixel was $35\arcsec\ \times 35\arcsec$. The data for all 12 epochs for each pointing were included in each image, since the much better \uv\ coverage (Fig.~\ref{fig:uvcover}) provides for much better image fidelity. Not all pointings had good data at all epochs (for example, because of a failed observation, or severe corruption by RFI).\label{sec:corruptreg} Some had bad data at all epochs because they were close to very bright sources -- in particular, in the regions surrounding 3C\,286 and 3C\,295. We included only pointings for which at least 9 out of 12 epochs were good -- a total of 243 pointings. The majority of these (218) had good data for all 12 epochs. The coverage of the final mosaic is shown in Fig.~\ref{fig:deepfield}.

\begin{figure}[ht]
\centering
\includegraphics[width=\linewidth,draft=false,bb=106 320 483 478]{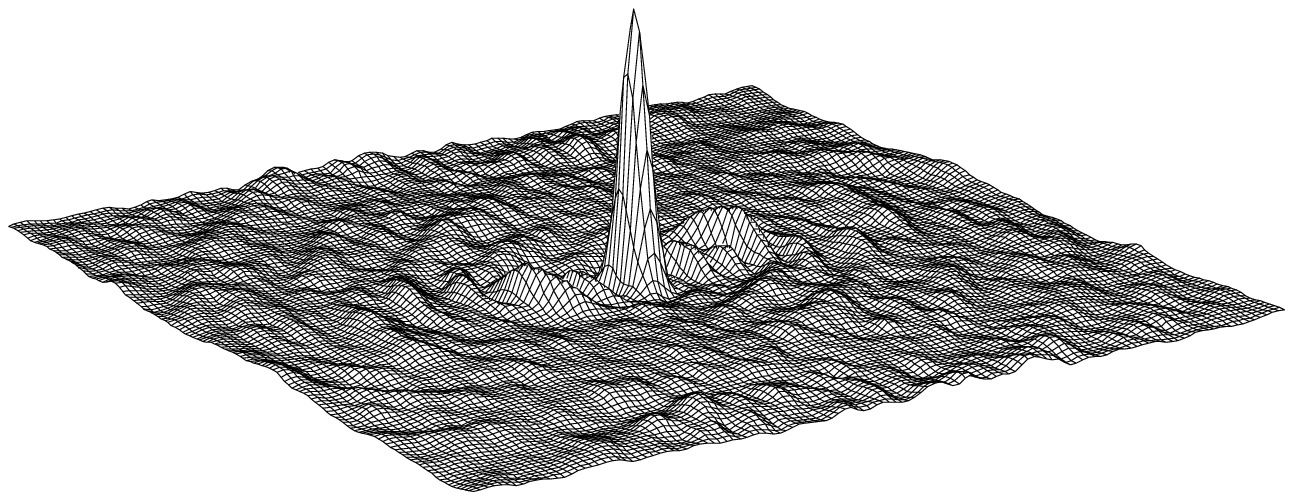}
\caption{\label{fig:beamplot}
A typical synthesized beam for ATATS, using data for all 12 epochs for a typical pointing (Field 192). The portion of the beam shown here is $75\arcmin \times 75\arcmin$. The sidelobes are $<10$\%\ of the intensity of the main lobe. Each field was CLEANed with its corresponding synthesized beam, but a common Gaussian restoring beam was used in producing the CLEANed images.
}
\end{figure}

\begin{figure}
\centering
\includegraphics[width=\linewidth,draft=false]{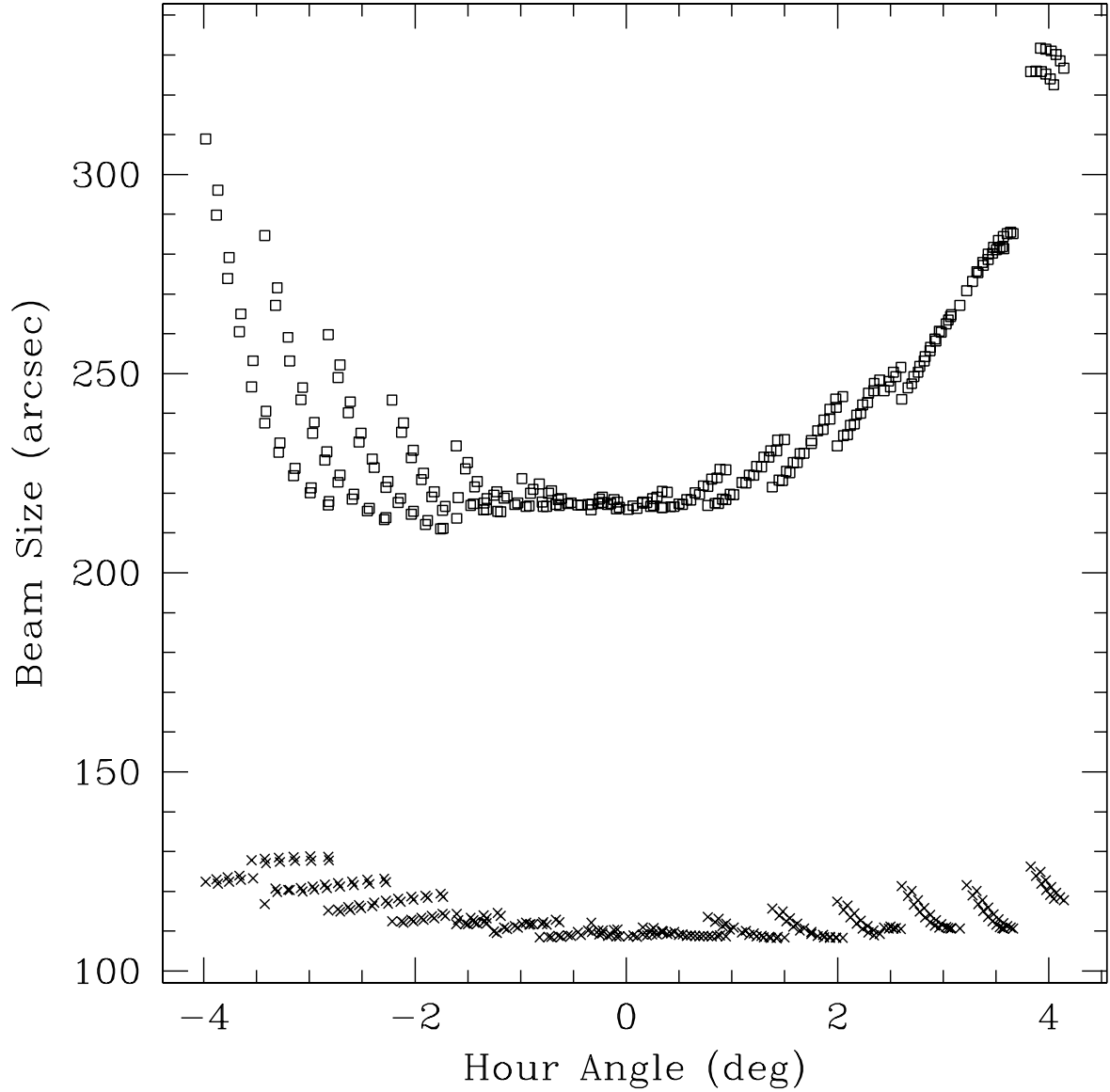}
\caption{\label{fig:beamsize}
The major axis width (squares) and minor axis width (crosses) of the ATATS synthesized beam as a function of hour angle, for the observations of 2009 March 31.
}
\end{figure}

The images were all made with the same sky coordinate for the reference pixel, to ensure that the resulting mosaic should have the correct projection geometry. A typical synthesized beam is shown in Fig.~\ref{fig:beamplot}. Since the synthesized beam changes in shape and position angle across the mosaic, it was necessary to use an ``average'' restoring beam to avoid superposing different beams from neighboring pointings, which creates problems at the catalog creation stage. We examined the beam major and minor axis width as a function of hour angle (Fig.~\ref{fig:beamsize}), and took the geometric mean of the major and minor axis width at ${\rm HA} = 0$, 150\arcsec, as the FWHM of a circular Gaussian which we used as a restoring beam. 

\subsection{Primary Beam Correction}

The ATA primary beam has been measured by various methods, including by off-axis measurement of point sources \citep[a method known as ``hex-7'';]{hex7}; by radio holography \citep{harp}; and by comparison of the flux densities of sources visible in two or more ATATS mosaic pointings \citep{hull}. These measurements give consistent results for the FWHM at 1.4\,GHz of 150\arcmin. We approximated the primary beam by a Gaussian model. We also experimented with varying the primary beam FWHM value used for mosaicking over a range of values, and found that the canonical value of 150\arcmin\ gave us the most accurate and precise fluxes for NVSS sources in our data, and so this value was adopted for our analysis. We applied the primary beam correction to the images of each good pointing.

Where the primary beam is well modeled, and where data volumes are small, joint deconvolution (where the entire mosaic is imaged and CLEANed at once) is generally desirable to maximize image fidelity. ATATS uses snapshot observations (where \uv\ coverage is not as good), and the primary beam model used is not as sophisticated, so better quality images can be obtained by imaging of individual fields, and subsequent mosaicking. In any case, the large amount of ATATS data would make joint deconvolution of the entire mosaic difficult without some averaging of the input visibilities (which risks degradation of image quality).

\subsection{Mosaicking}

We made a linear mosaic of the 243 individual primary-beam corrected images, masking each outside a radius corresponding to the nominal half-power points (75\arcmin), in order to avoid exaggerating the effects of the deviation of the true primary beam from a Gaussian at large distances from the pointing center \citep[in particular, the $10\%$ sidelobes;][]{harp}.

The final image was transformed to a GLS (global sinusoid) geometry, which has the property that each pixel subtends an equal area on the sky. This is important for an image covering such a large area, since the curvature of the sky introduces distortions. In projections that do not preserve area, the size of the beam would change across the mosaic, resulting in errors in the integrated flux densities reported by the source fitting procedure. If we were not to perform this transformation on the ATATS mosaic, we would bias the flux densities of the sources low by around $\sim 2\%$, and increase the scatter in the measured flux densities by $\sim 12\%$ when compared to NVSS.

\begin{figure}
\centering
\includegraphics[angle=270,width=\linewidth,draft=false]{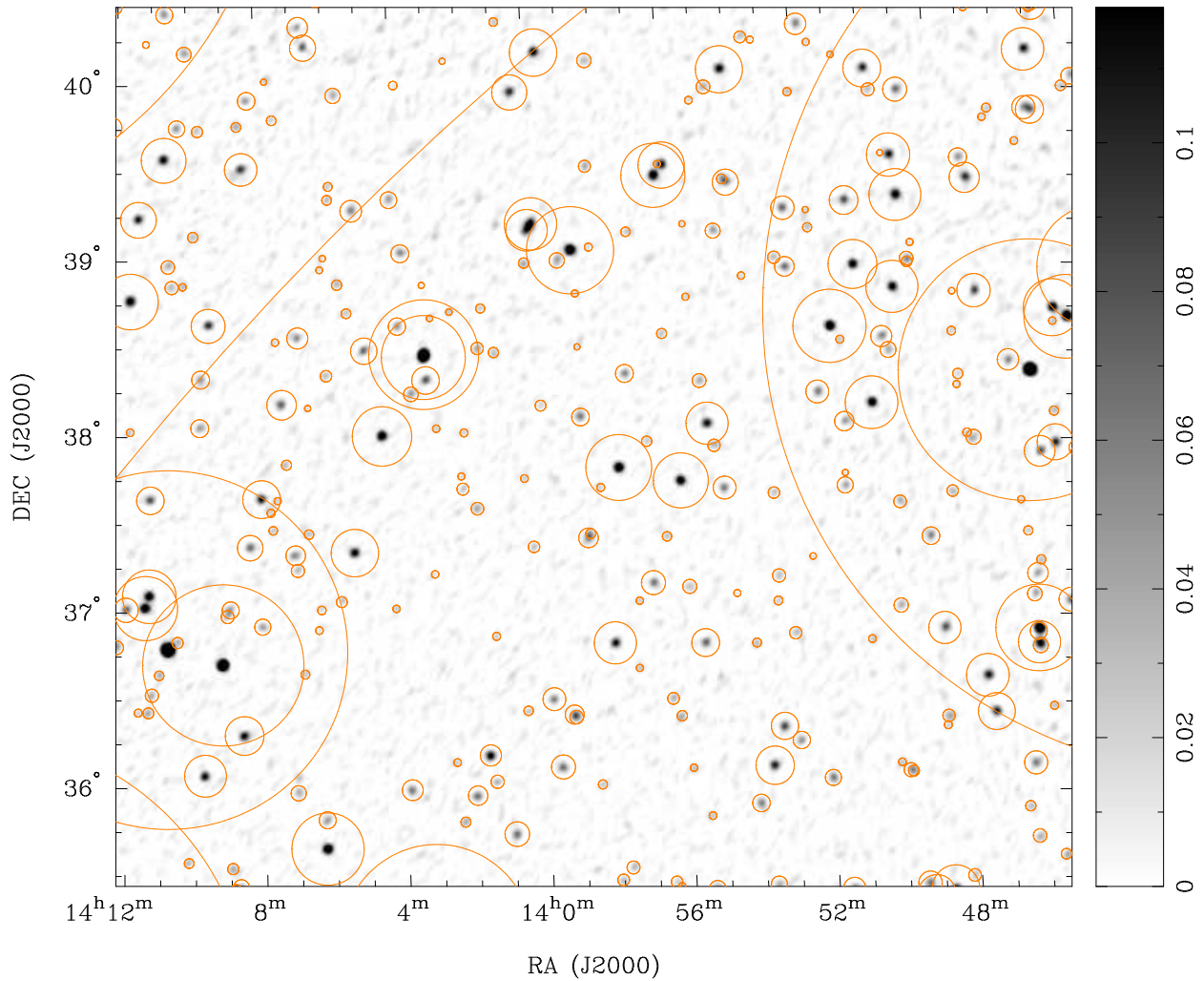}
\caption{\label{fig:deepfieldzoom}
A 25~\sqdeg\ region of the deep field shown in Fig.~\ref{fig:deepfield}, shown at a larger scale and slightly different stretch so that the structure of individual sources may be seen. All sources with NVSS flux densities brighter than 20\,mJy (corresponding to 5 times the RMS of the ATATS image) are plotted as circles; the size of the circle is proportional to the NVSS flux density. The greyscale runs from zero to 118.2\,\mjypbm ($30 \sigma$). The faintest NVSS sources are below the ATATS completeness limit. Hence the fraction of NVSS sources without a counterpart in the ATATS catalog increases with decreasing circle size in this image.
}
\end{figure}

The RMS noise in this final 12-epoch image, which we term the ``master mosaic'', is 3.94\,\mjypbm. The dynamic range, determined by dividing the peak flux density of the brighest source in our map (4.14\,Jy) by the RMS noise in a 100-pixel box surrounding this source (23.0\,mJy) is 180. The master mosaic covers 689.63~\sqdegs\ of sky -- hence, the mean master mosaic pixel has contributions from 1.73 pixels of the 243 input 12-epoch images, each of which covers 4.91~\sqdegs. The master mosaic is shown in Fig.~\ref{fig:deepfield}, and a smaller region, which better shows the individual ATATS sources, is shown in Fig.~\ref{fig:deepfieldzoom}.

\subsection{Catalog creation}

The MIRIAD task SFIND was run on the master mosaic. The RMS was computed over a box of side 100 pixels, and sources above 5 times the local RMS were kept.\label{sec:sfind} The resulting SFIND catalog contained 4843 sources. Of these, 197 were poorly fit by SFIND (indicated by asterisks in the SFIND output) and were rejected from the catalog, leaving 4646 sources. Of these, 238 sources were in regions where 10\%\ or more of the pixels in a $1\degr \times 1\degr$ box centered on the source were masked (\S~\ref{sec:mask}). These sources were also rejected, leaving 4408 sources in the final catalog (Table~\ref{tab:catalog}).

\begin{deluxetable}{rrrrrrrrr}
\tablewidth{0pt}
\tabletypesize{\scriptsize}
\tablecaption{\label{tab:catalog} The ATATS catalog}
\tablehead {
\colhead{RA (J2000)} &
\colhead{Dec (J2000)} &
\colhead{\satats\ (mJy)} &
\colhead{$\sigma_{\rm bg}$(mJy)} &
\colhead{$\Theta_{\rm maj}$ (\arcsec)} & 
\colhead{$\Theta_{\rm min}$ (\arcsec)} &
\colhead{PA (\degr)} &
\colhead{N$_{\rm NVSS}$ \tablenotemark{a}} &
\colhead{\snvss\ (mJy) \tablenotemark{b}}
}
\startdata
\hms{12}{55}{50.4} & \dms{53}{03}{51} & 85.2 & 4.9 & 164.4 & 131.1 & 41.8 & 1 & 79.3\\
\hms{12}{55}{52.3} & \dms{52}{49}{59} & 102.4 & 4.1 & 177.4 & 137.1 & 45.0 & 1 & 98.1\\
\hms{12}{57}{11.2} & \dms{53}{35}{11} & 25.1 & 5.1 & 164.6 & 115.7 & 28.3 & 1 & 39.4\\
\hms{12}{57}{11.6} & \dms{52}{43}{24} & 37.8 & 4.2 & 204.7 & 138.4 & 7.7 & 1 & 34.2\\
\hms{12}{57}{37.4} & \dms{52}{38}{26} & 105.8 & 5.6 & 175.1 & 136.3 & 24.5 & 1 & 102.4\\
\hms{12}{58}{00.1} & \dms{52}{22}{51} & 49.8 & 5.4 & 168.8 & 124.8 & 11.9 & 1 & 59.8\\
\hms{12}{58}{02.2} & \dms{52}{51}{47} & 80.0 & 4.2 & 194.4 & 137.8 & 61.9 & 1 & 70.2\\
\hms{12}{58}{14.1} & \dms{54}{21}{53} & 678.6 & 7.3 & 173.9 & 142.0 & 42.7 & 1 & 647.0\\
\hms{12}{58}{42.5} & \dms{54}{43}{38} & 60.6 & 4.4 & 166.9 & 133.8 & 44.0 & 1 & 58.7\\
\hms{12}{59}{18.6} & \dms{54}{32}{21} & 95.0 & 4.9 & 190.2 & 137.7 & 44.9 & 1 & 82.0\\
\hms{12}{59}{37.8} & \dms{53}{20}{44} & 40.4 & 3.5 & 180.2 & 167.0 & -29.8 & 1 & 33.3\\
\hms{13}{00}{23.0} & \dms{52}{35}{39} & 67.6 & 4.7 & 170.0 & 136.2 & 46.1 & 1 & 71.9\\
\hms{13}{00}{38.2} & \dms{53}{11}{49} & 37.5 & 5.1 & 210.2 & 155.0 & 61.2 & 1 & 18.1\\
\hms{13}{01}{12.0} & \dms{53}{14}{22} & 35.6 & 5.4 & 177.0 & 119.3 & -0.9 & 1 & 40.2\\
\hms{13}{01}{22.1} & \dms{53}{32}{56} & 120.3 & 3.4 & 164.2 & 141.0 & 45.2 & 1 & 115.5\\
\hms{13}{01}{40.5} & \dms{54}{08}{28} & 301.3 & 3.6 & 248.8 & 140.9 & 65.9 & 1 & 167.9\\
\hms{13}{01}{43.5} & \dms{52}{53}{58} & 54.3 & 3.3 & 257.1 & 193.9 & -43.1 & 1 & 22.9\\
\hms{13}{02}{28.1} & \dms{54}{13}{52} & 52.1 & 4.4 & 168.8 & 136.5 & 26.3 & 1 & 54.4\\
\hms{13}{02}{32.5} & \dms{54}{51}{41} & 58.0 & 4.4 & 169.9 & 122.6 & 23.6 & 1 & 60.8\\
\hms{13}{02}{45.9} & \dms{53}{16}{58} & 44.0 & 3.9 & 204.5 & 139.8 & 12.4 & 1 & 41.1\\
\hms{13}{02}{59.7} & \dms{52}{40}{36} & 41.8 & 4.2 & 223.5 & 129.9 & 47.5 & 1 & 37.0\\
\hms{13}{03}{11.3} & \dms{54}{42}{22} & 28.9 & 5.1 & 171.6 & 124.5 & 31.5 & 1 & 31.5\\
\hms{13}{03}{15.0} & \dms{49}{16}{02} & 105.1 & 4.3 & 185.4 & 140.4 & 24.6 & 1 & 90.1\\
\hms{13}{03}{26.9} & \dms{49}{35}{51} & 41.1 & 3.8 & 203.7 & 147.0 & -57.9 & 1 & 48.1\\
\hms{13}{03}{28.3} & \dms{54}{24}{41} & 77.3 & 3.7 & 189.2 & 134.0 & 18.2 & 1 & 70.7\\
\hms{13}{03}{30.1} & \dms{53}{43}{53} & 48.0 & 3.7 & 169.3 & 143.8 & 22.3 & 1 & 47.9\\
\hms{13}{03}{57.9} & \dms{52}{46}{27} & 57.3 & 4.5 & 152.9 & 133.0 & 11.1 & 1 & 69.3\\
\hms{13}{04}{03.8} & \dms{51}{28}{12} & 69.3 & 5.0 & 208.2 & 140.9 & 46.1 & 1 & 68.3\\
\hms{13}{04}{17.9} & \dms{50}{24}{09} & 30.6 & 3.7 & 167.8 & 127.9 & -9.4 & 1 & 33.4\\
\hms{13}{04}{27.4} & \dms{53}{50}{06} & 916.5 & 6.4 & 173.7 & 147.1 & 50.8 & 2 & 924.9\\
\hms{13}{04}{27.9} & \dms{51}{43}{25} & 81.9 & 5.4 & 182.4 & 155.6 & 40.6 & 1 & 74.2\\
\hms{13}{04}{35.6} & \dms{49}{54}{15} & 35.4 & 3.9 & 195.3 & 149.3 & -6.9 & 1 & 30.2\\
\hms{13}{04}{36.5} & \dms{54}{28}{36} & 25.1 & 4.1 & 179.5 & 129.4 & 40.7 & 1 & 28.9\\
\hms{13}{04}{42.1} & \dms{53}{24}{21} & 213.0 & 5.6 & 192.6 & 146.0 & 61.1 & 1 & 208.1\\
\hms{13}{04}{43.3} & \dms{50}{12}{46} & 31.8 & 5.0 & 157.6 & 139.1 & 17.2 & 1 & 36.1\\
\hms{13}{04}{45.6} & \dms{50}{56}{26} & 343.1 & 5.4 & 170.4 & 138.8 & 45.0 & 1 & 345.7\\
\hms{13}{04}{57.7} & \dms{53}{59}{16} & 86.6 & 3.8 & 167.4 & 141.3 & 37.5 & 1 & 87.6\\
\hms{13}{05}{03.3} & \dms{51}{07}{31} & 49.0 & 5.4 & 179.1 & 147.6 & -7.5 & 1 & 56.7\\
\hms{13}{05}{06.0} & \dms{52}{14}{32} & 44.3 & 3.4 & 171.1 & 153.4 & 29.4 & 1 & 43.3\\
\hms{13}{05}{06.8} & \dms{52}{53}{16} & 57.3 & 4.2 & 173.1 & 152.0 & 24.9 & 1 & 79.9\\
\enddata
\tablenotetext{a}{Number of NVSS matches within 75\arcsec of ATATS position}
\tablenotetext{b}{Sum of integrated flux densities of NVSS sources within 75\arcsec of ATATS position}
\tablecomments{Only a portion of this table is shown here to demonstrate its form and content. A machine-readable version of the full table is available.}
\end{deluxetable}

\section{Analysis}

In order to analyze the quality and reliability of our data, we compared the images and catalogs to those produced by the NVSS \citep{nvss}, a survey at 1.4\,GHz (with 45\arcsec\ resolution) undertaken with the VLA between 1993 and 1997. One of the goals of ATATS is to study source variability; we expect some sources to have changed in brightness, perhaps by so much that sources in NVSS may not be present in ATATS or vice versa. However, the majority of sources in the NVSS should have similar fluxes and positions when measured again by ATATS, and by looking at variations in these quantities we can assess the quality of the ATATS data. We can also compare fluxes and positions from epoch to epoch in ATATS; variability on epoch-to-epoch timescales will be discussed by \citet{paperii}.

\begin{figure}
\centering
\includegraphics[width=\linewidth,draft=false]{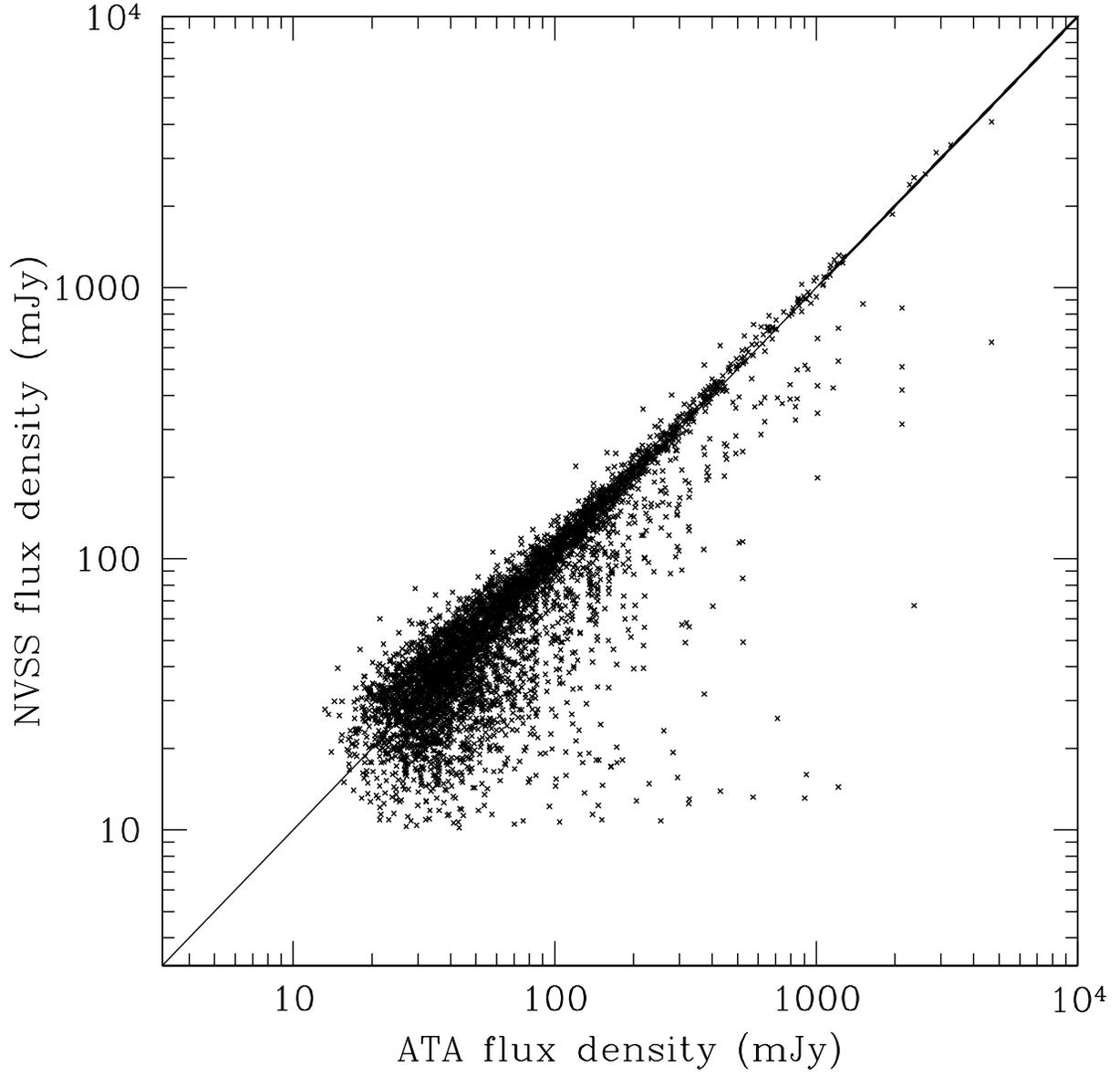}%
\caption{\label{fig:fluxmatch}
Comparison of the flux densities of the closest ATATS source (if one exists within a radius of 75\arcsec) to each NVSS catalog position.
}
\end{figure}

We matched the ATATS master mosaic catalog with NVSS in a number of ways, experimenting with different source-finding parameters, match radii, and rejection of spurious sources in order to obtain the most reliable catalog. We found that a simple circular positional match with a match radius of 75\arcsec\ (corresponding to the size of the restoring beam) gave a good compromise between too many false matches, and too many true matches that might be missed. However, it should be noted that the differing resolutions and sensitivities of ATATS and NVSS mean that the ``correct'' NVSS source is not always matched to the corresponding ATATS source. In order to reduce the number of spurious matches of ATATS sources to NVSS, we consider only NVSS sources brighter than 10\,mJy, around 2.5 times the RMS of the ATATS master mosaic. We search the ATATS SFIND catalog around the positions of NVSS sources, and plot the resulting flux densities of matched sources in Fig.~\ref{fig:fluxmatch}.

Sometimes a source unresolved by ATATS is resolved by NVSS into two or more components. This results in apparent changes in flux density for the matched catalog sources, since only one of the components (and hence only a part of the object's true flux) at a time is matched at these epochs. Because we search around each NVSS source for the nearest ATATS source, there are instances where different NVSS sources are each matched to the same nearby ATATS source. These cases appear as vertically aligned distributions of points in Fig.~\ref{fig:fluxmatch}, and are more common at faint fluxes, where the areal density of sources is higher. There are also cases where a faint NVSS source matches a bright ATA source because the ATA either does not detect the fainter source (and so the NVSS source is matched to a nearby brighter, unassociated ATATS source), or because the larger synthesized beam of the ATA results in the detection of a single source which NVSS resolves into two or more components. In such cases the resulting ATATS flux density is approximately the sum of the flux densities of the NVSS components. There may also be cases where the ATA detects extended emission which is resolved out by the higher-resolution NVSS.

\begin{figure}
\centering
\includegraphics[width=\linewidth,draft=false]{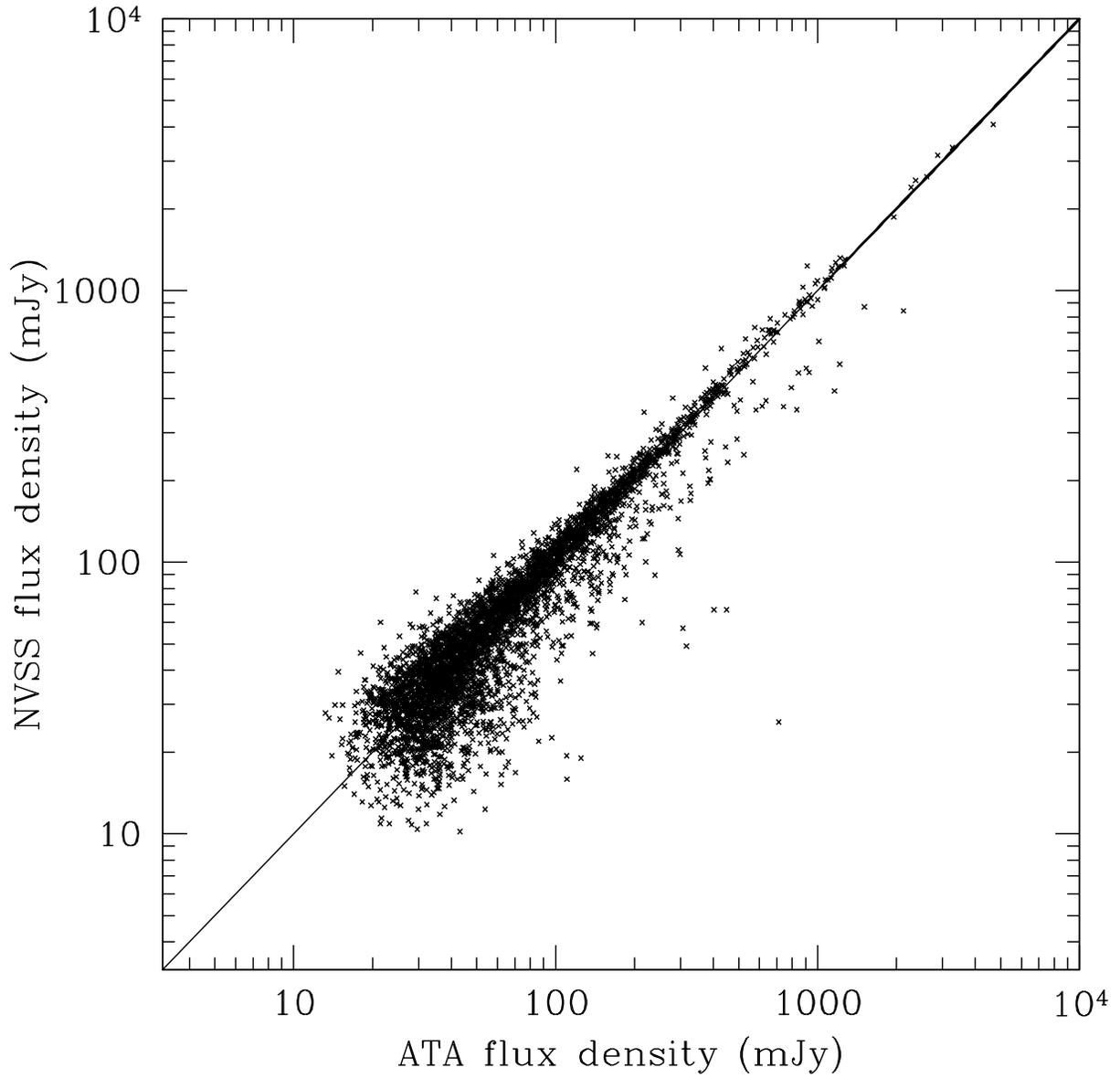}
\caption{\label{fig:fluxmatchatats}
Comparison of the flux densities of the closest NVSS source (if one exists within a radius of 75\arcsec) to each ATATS catalog position. Note that the sense of the match is reversed compared to Fig.~\ref{fig:fluxmatch}.
}
\end{figure}

Because the flux limit of the NVSS catalog is fainter than that for ATATS, the number of NVSS sources within the ATATS field is larger than the number of ATATS sources in the same region, and so we can ameliorate the problem of multiple matches of the same ATATS source to several NVSS sources by reversing the sense of the match. We searched the ATATS catalog for the closest NVSS match, again with a 75\arcsec\ match radius, and plotted the matched sources in Fig.~\ref{fig:fluxmatchatats}). The ``tail'' of sources to the lower right of Fig.~\ref{fig:fluxmatch}, where a bright ATATS source is matched several times to the ``wrong'' nearby NVSS source, is largely gone.

However, there are still a number of sources far below the 1:1 line, and fewer sources far above. It is equally likely that a source becomes ten times brighter between the NVSS and ATATS epochs, as that a source becomes ten times fainter. Were both datasets taken with the same instrument, and to the same flux density limit, the matched catalogs would be expected to produce a symmetrical cloud of points around the 1:1 line, with the scatter due to a combination of measurement uncertainty and intrinsic variability. The slight asymmetry of the point cloud in Fig.~\ref{fig:fluxmatchatats} is due to the different resolutions and sensitivities of NVSS and ATATS.\label{sec:symmetric} To attempt to account for this, and more reliably compare the two catalogs, we ``smoothed'' the NVSS catalog to match the ATATS resolution, by summing the fluxes of all NVSS sources within 75\arcsec\ of the ATATS position. The number of NVSS sources matching each ATATS source, and their summed flux density, are shown in Table~\ref{tab:catalog}.\label{sec:smooth}

\begin{figure}
\centering
\includegraphics[width=\linewidth,draft=false]{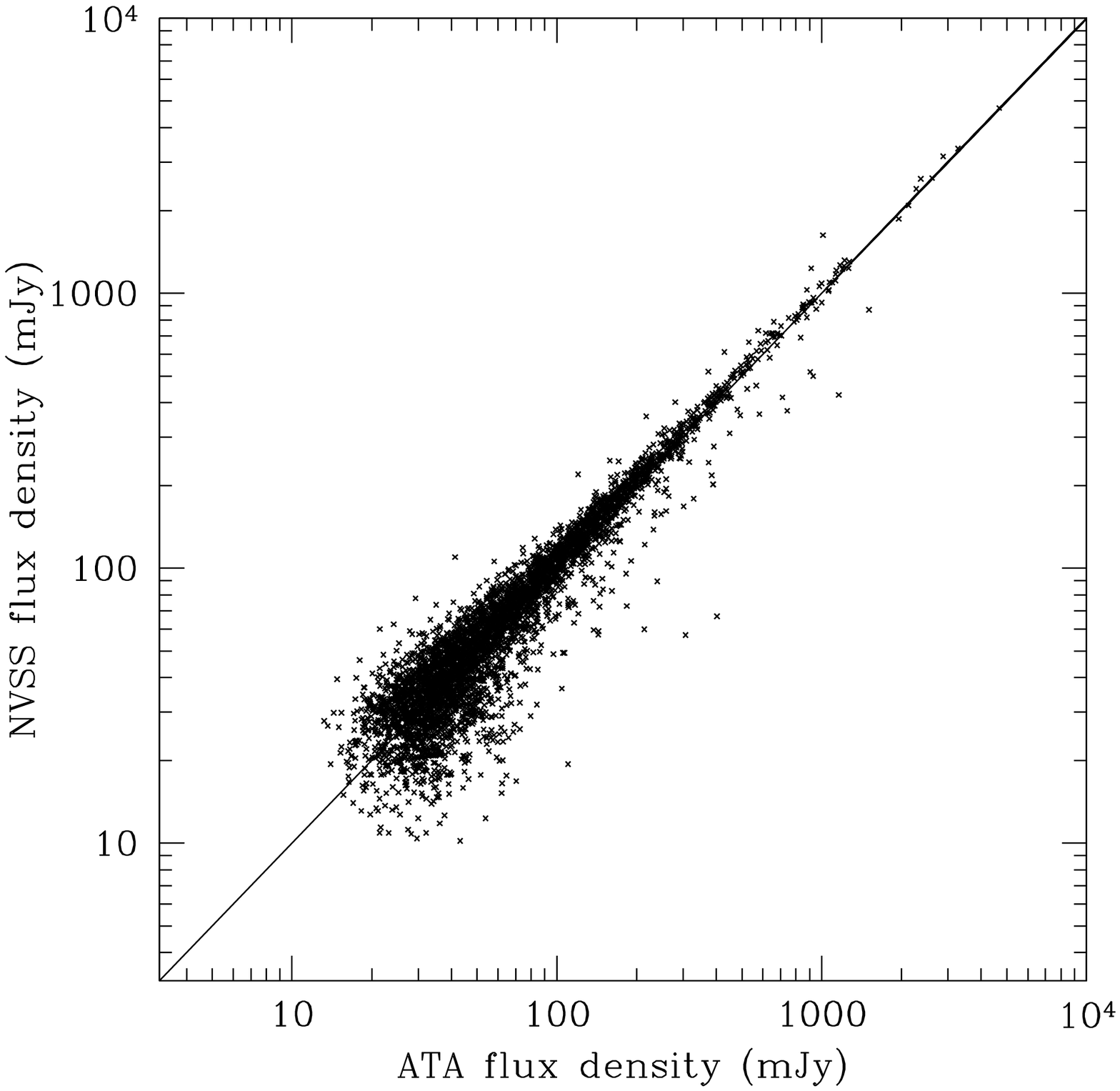}
\caption{\label{fig:fluxmatchsmooth}
Comparison of the flux densities of sources from the ATATS catalog, with the sum of fluxes of all NVSS sources within 75\arcsec\ of the ATATS positions.
}
\end{figure}

The effects of the smoothing can be seen in Fig.~\ref{fig:fluxmatchsmooth}. The summation of the fluxes of NVSS sources within the ATA synthesized beam brings the fluxes more closely into line with those from NVSS. However, it is clear that there are still some residual resolution effects, as evidenced by the fact that the point cloud is still somewhat asymmetric.

The spread of the points around the 1:1 line increases towards lower flux densities. This is mainly because percentage uncertainties in the flux measurement increase towards lower fluxes, but also partly because of intrinsic variability in the sources themselves over the $\sim 15$ years between the NVSS and ATATS observations (\S~\ref{sec:variability}). 

A least squares fit to the data in Fig.~\ref{fig:fluxmatchsmooth} has a gradient of $1.006 \pm 0.003$ and an intercept of $1.3 \pm 0.5$\,mJy, so the match of ATATS flux densities to NVSS is very good. In \S~\ref{sec:bias}, we also compare the median flux densities of ATATS and NVSS and find good agreement.

\subsection{Completeness}

By selecting only ATATS sources which match NVSS, we are able to estimate completeness of the ATATS catalogs; by selecting those which do not match an NVSS source, we are able to identify spurious sources and potential transients. Clearly we wish to reject spurious sources while keeping potential transients, and we tweaked the transient detection pipeline to achieve this. We found that the main cause of spurious sources was increased noise at the edges of mosaics, so when looking for transient candidates we reject those which are in fields with a significant fraction ($\geq 10\%$) of blanked or masked pixels in a $1\degr \times 1\degr$ box centered on the source. \label{sec:mask} 

\begin{figure}
\centering
\includegraphics[width=0.5\linewidth,draft=false]{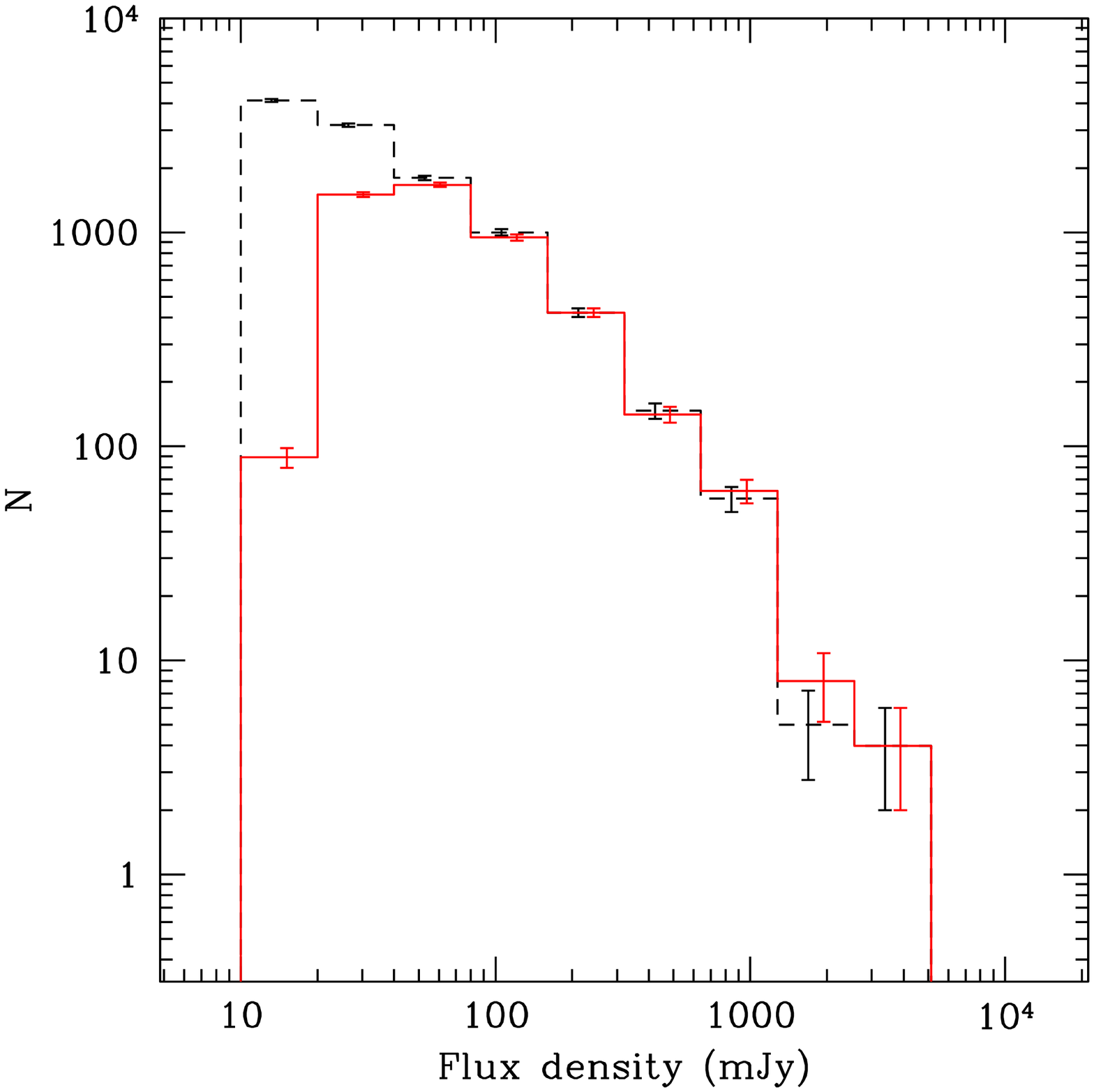}%
\includegraphics[width=0.5\linewidth,draft=false]{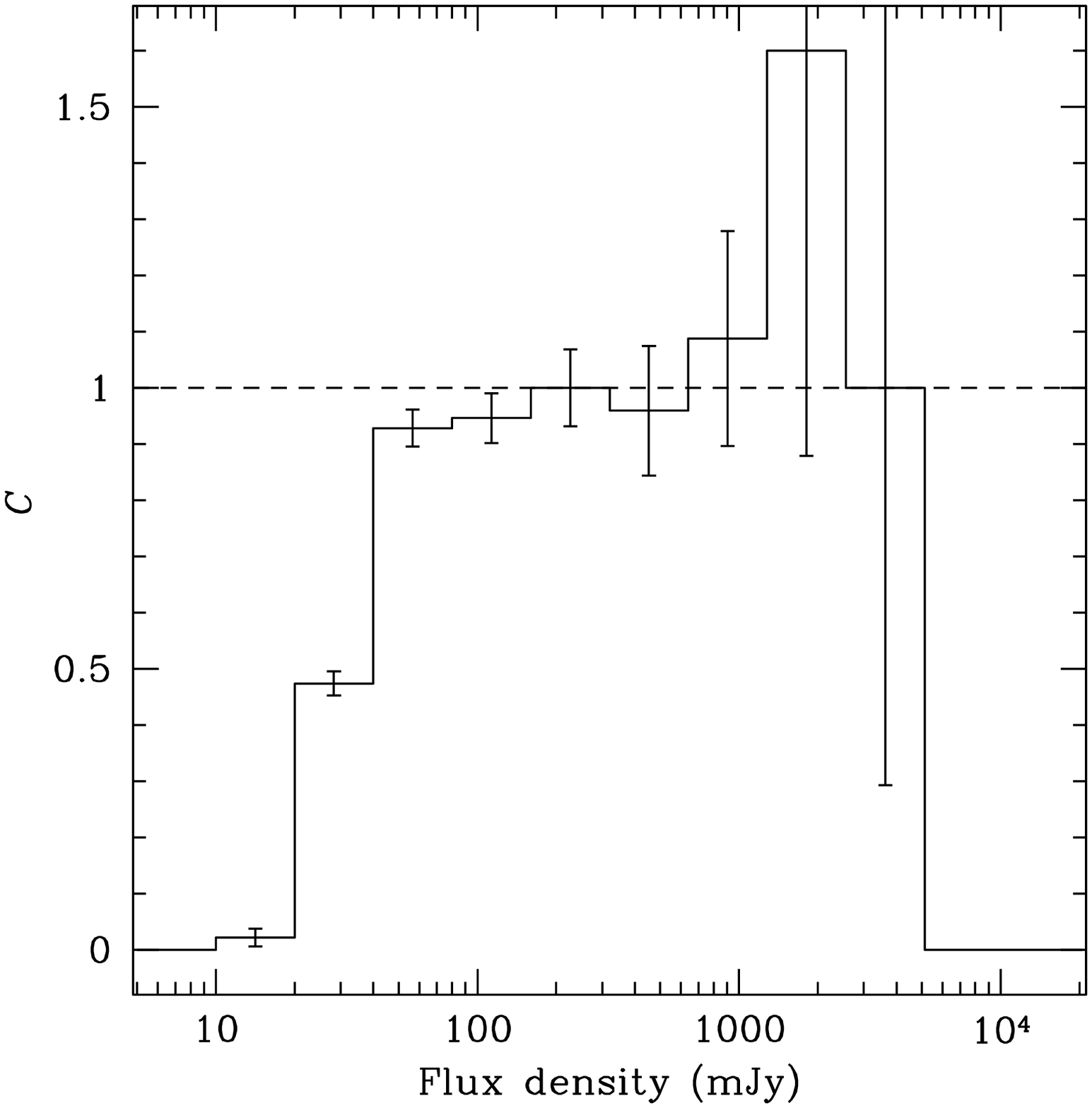}
\caption{\label{fig:compall}
{\em Left:} Log(N) -- log(S) histogram of NVSS sources (black dashed line) in good regions of the ATATS field (\ie, the culled catalogs -- see \S~\ref{sec:cull}), and sources detected in the ATATS catalog (red solid line) over the same area. Error bars assume Poisson statistics. Each bin covers a factor of 2 in integrated flux density, or 0.301 dex. The coarser resolution of the ATA tends to combine the flux from a tight cluster of NVSS sources into a single ATATS source. Combined with the ATA's additional sensitivity to structures larger than the NVSS synthesized beam, this tends to shift some sources from fainter into brighter bins. {\em Right:} \label{fig:complete}Ratio of the histograms for the ATATS and NVSS catalogs, $C = N_{\rm ATATS} / N_{\rm NVSS}$ . This represents a measure of the completeness of the ATATS catalogs. Error bars are computed by propagating the Poisson errors from the ATATS and NVSS source count histograms. The tendency of the ATA to shift sources into higher flux bins results in some bins with $C > 1$. These bins are consistent with 100\%\ completeness (dashed line) within the Poisson errors.
}
\end{figure}

Even without matching the catalogs source by source, we can assess the completeness and reliability of ATATS by plotting the flux density distribution of our catalogs and comparing to NVSS. In order to compare source counts over the same area, we culled\label{sec:cull} sources from the NVSS catalog in regions where ATATS has no data. The resulting culled NVSS catalog, which covers the same area as the ATATS master mosaic, contains 10729 sources down to a limiting flux density of 10\,mJy. We plot source count histograms of the ATATS and culled NVSS catalogs in Fig.~\ref{fig:compall}. The two histograms are consistent, within the errors, until the ATATS counts turn over between 40 and 80\,mJy.

The NVSS histogram truncates abruptly at 10\,mJy due to the cut we impose on the catalog. This is well above the 99\%\ completeness limit of the NVSS, 3.4\,mJy, so the NVSS counts may be considered a measure of the true source counts for this region of sky. The ATATS histogram is marginally higher than that for NVSS for the bins between 320 and 1280\,mJy, and marginally lower for the bins between 40 and 160\,mJy (above the points where the source counts turn over). This is to be expected since the poorer resolution of ATATS will tend to combine the flux of multi-component NVSS sources into a single brighter ATATS source, and occasionally to detect extended flux resolved out by NVSS. This should only affect a small number of sources though, so the effect on the histogram is not large.

The ratio of the ATATS to NVSS histograms can also be interpreted as the efficiency with which we would detect transient sources of a given brightness, or as a measure of the survey completeness. We plot the ratio of the two histograms, which we denote $C$, in Fig.~\ref{fig:complete}. We can see that ATATS is $> 90\%$ complete down to around 40\,mJy, or approximately 10 times the RMS noise in the master mosaic, below which the completeness falls off rapidly. The shift of some sources into higher flux bins due to the resolution mismatch results in some bins with computed $C > 1$. So some of the sources ``missing'' at $S \gtrsim 40$\,mJy are in fact detected, but with higher flux density, and the 90\%\ completeness value may be considered a lower limit. Matching of individual ATATS sources to NVSS and subsequent examination of postage stamp images (\S~\ref{sec:99.9complete}) also supports the conclusion that 90\%\ is a conservative value. 

As shown in Fig.~\ref{fig:deepfieldzoom}, which plots a region of the master mosaic, with NVSS sources brighter than 20\,mJy overlaid, there is an excellent match between the two surveys, although counterparts to some of the faintest sources above the NVSS cut are sometimes marginal or non-detections in ATATS.

\begin{figure}
\centering
\includegraphics[width=\linewidth,draft=false]{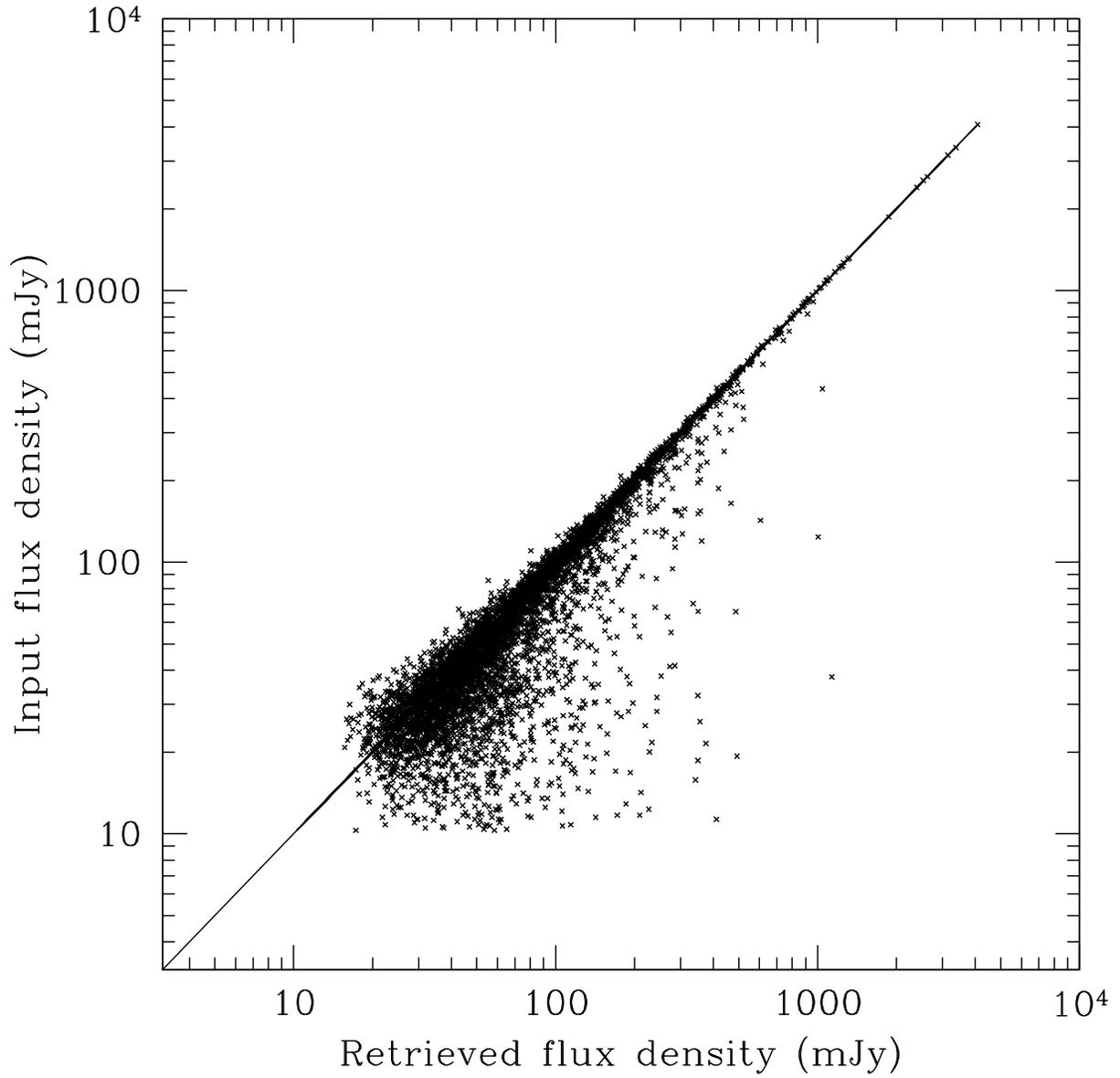}
\caption{\label{fig:fluxmatchfake}
Comparison of the flux densities of simulated sources inserted into the ATATS master mosaic, and the fluxes retrieved by SFIND for the nearest source within 75\arcsec\ of the input positions.
}
\end{figure}

We are also able to estimate completeness by adding simulated sources to the ATATS mosaic. We took the flux densities of the 10729 sources in the culled NVSS catalog, assigned them to random positions within the ATATS mosaic, and created simulated Gaussian sources with FWHM 150\arcsec\ at these positions. We then ran SFIND using the same parameters as used for the generation of the science catalog (\S~\ref{sec:sfind}). The resultant catalog contains both the real sources detected in ATATS, and those of the simulated sources that were recovered by SFIND. In order to determine our ability to accurately recover the input flux densities, we searched for the closest ATATS source (within a maximum radius of 75\arcsec) to the input positions of the simulated NVSS sources. The resulting flux--flux diagram is shown in Fig.~\ref{fig:fluxmatchfake}. This exercise is similar to the matching of the real NVSS catalog with the ATATS catalog, as seen in Fig.~\ref{fig:fluxmatch}, modulo the fact that adding simulated sources increases the number of detections in the SFIND catalog from 4843 to 9404 (which will increase source confusion somewhat), that resolution effects apply to the match with NVSS, but not to the retrieval of simulated sources, and that intrinsic source variability will result in changed fluxes in Fig.~\ref{fig:fluxmatch}, whereas no such effect should be seen in Fig.~\ref{fig:fluxmatchfake}.\label{sec:realvfake}

We also searched for ATATS sources within 75\arcsec\ of NVSS sources within the region where we have good ATATS data. This provides a similar (and consistent) estimate of completeness to the ratio of the ATATS and NVSS histograms in Fig.~\ref{fig:complete}. 93\%\ of NVSS sources brighter than 40\,mJy are detected by ATATS, as are 97\%\ of sources brighter than 120\,mJy. \label{sec:nomatch} The remaining 3\%\ of sources without a match in the ATATS catalog are in fact clearly visible in the mosaic images, but are not present in the ATATS catalog due to poor fits or failed deblending (\S~\ref{sec:nobrighttransients}).

\subsection{Positional Accuracy}

\begin{figure}
\centering
\includegraphics[width=0.5\linewidth,draft=false]{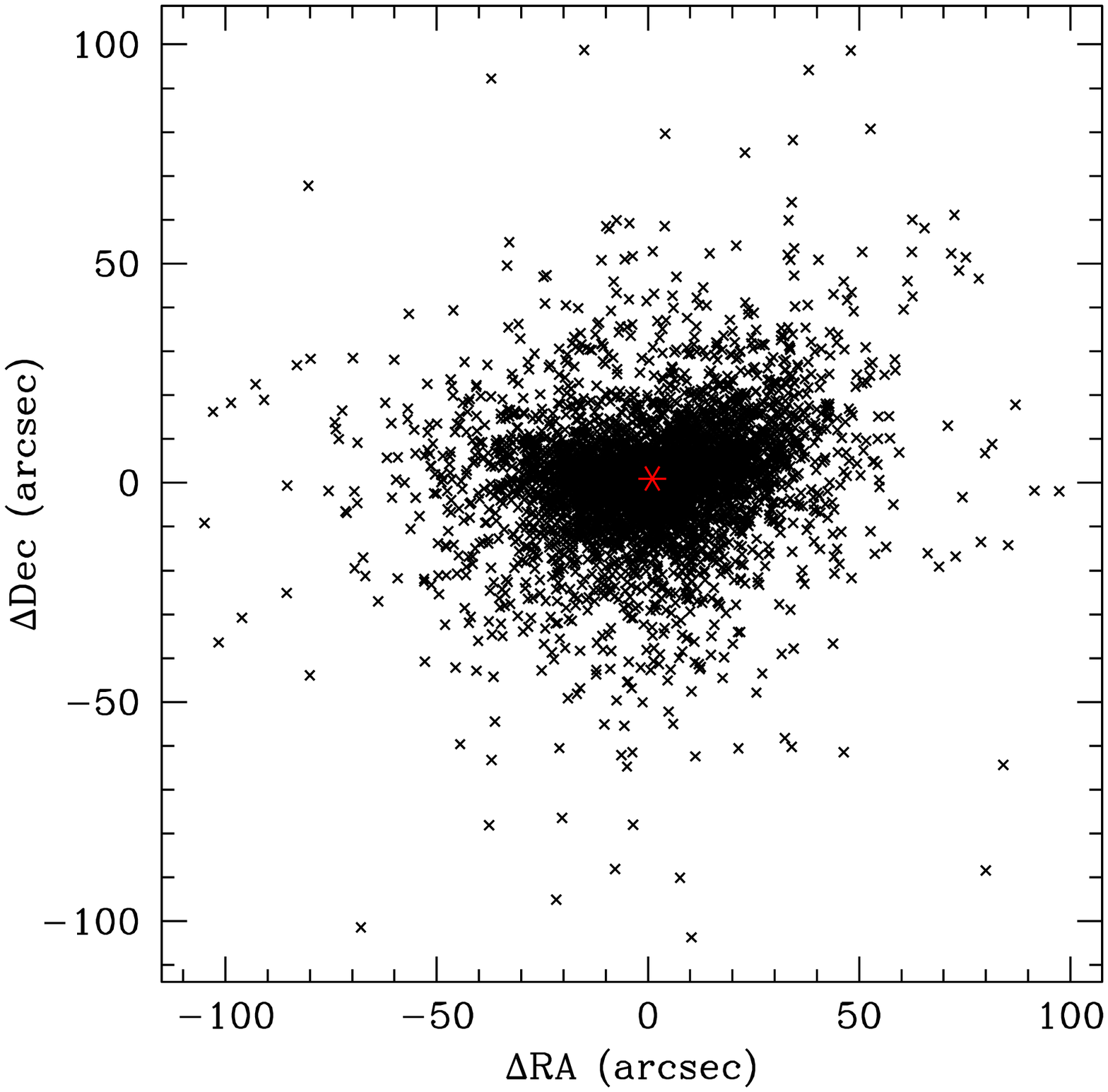}%
\includegraphics[width=0.5\linewidth,draft=false]{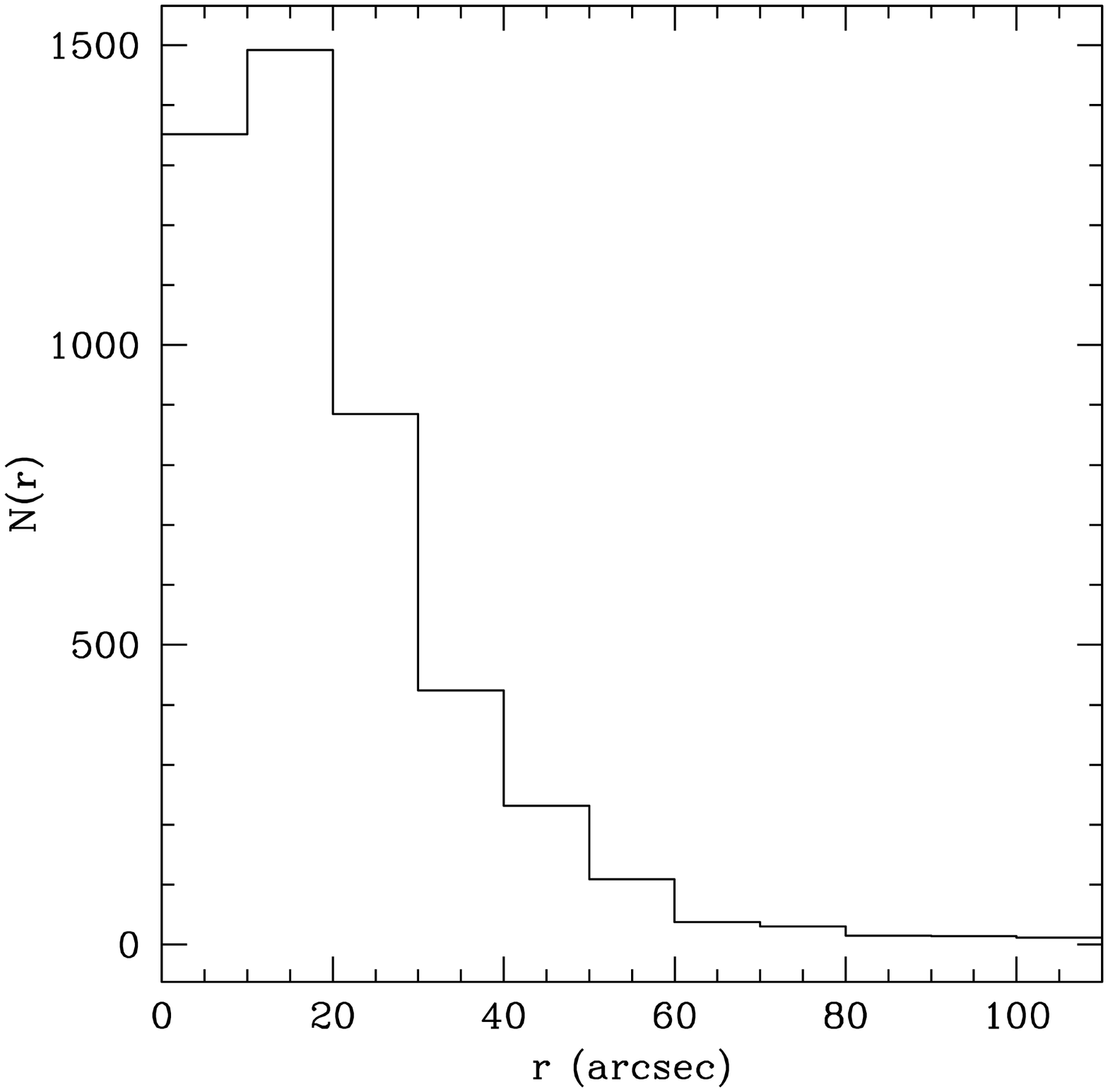}
\caption{\label{fig:offsets}
{\em Left:} Positional offsets in RA and Dec for sources detected by both NVSS and ATATS.
{\em Right:} Histogram of offsets ($\sqrt{\Delta{\rm RA}^2 + \Delta{\rm Dec}^2}$) for sources detected by both NVSS and ATATS. The majority of the matches fall within a radius corresponding to the ATA synthesized beam ($\sim 75$\arcsec) -- those at larger radii are mostly spurious matches of an NVSS source with an unassociated nearby ATA source, where the true match was not detected by the ATA. 
}
\end{figure}

To determine the positional accuracy of ATATS versus NVSS (which has an intrinsic positional accuracy typically $\lesssim 3$\arcsec) we plot the offsets in right ascension and declination, and a histogram of the magnitude of the offset between ATATS and NVSS positions (Fig.~\ref{fig:offsets}). A 120\arcsec\ match radius was used for these plots, but as can be seen, the majority of the matches fall within a radius corresponding to the ATA synthesized beam ($\sim 75$\arcsec), the value adopted for the match radius in our analysis.

The median offset of the ATATS position from the NVSS position is ($\Delta{\rm RA}, \Delta{\rm Dec}) = (1\farcs4,0\farcs98)$, and the RMS scatter in the positional offsets is ($\sigma_{\rm RA}, \sigma_{\rm Dec}) = (20\farcs4,14\farcs7)$.

As shown in Fig.~\ref{fig:deepfieldzoom}, the positional accuracy is sufficient in most cases to provide a clear match between ATATS and NVSS sources, although in some cases a single ATATS source is associated with multiple NVSS sources due to the differing resolutions of the two surveys.

\subsection{Variability\label{sec:variability}}

\begin{figure}
\centering
\includegraphics[width=\linewidth,draft=false]{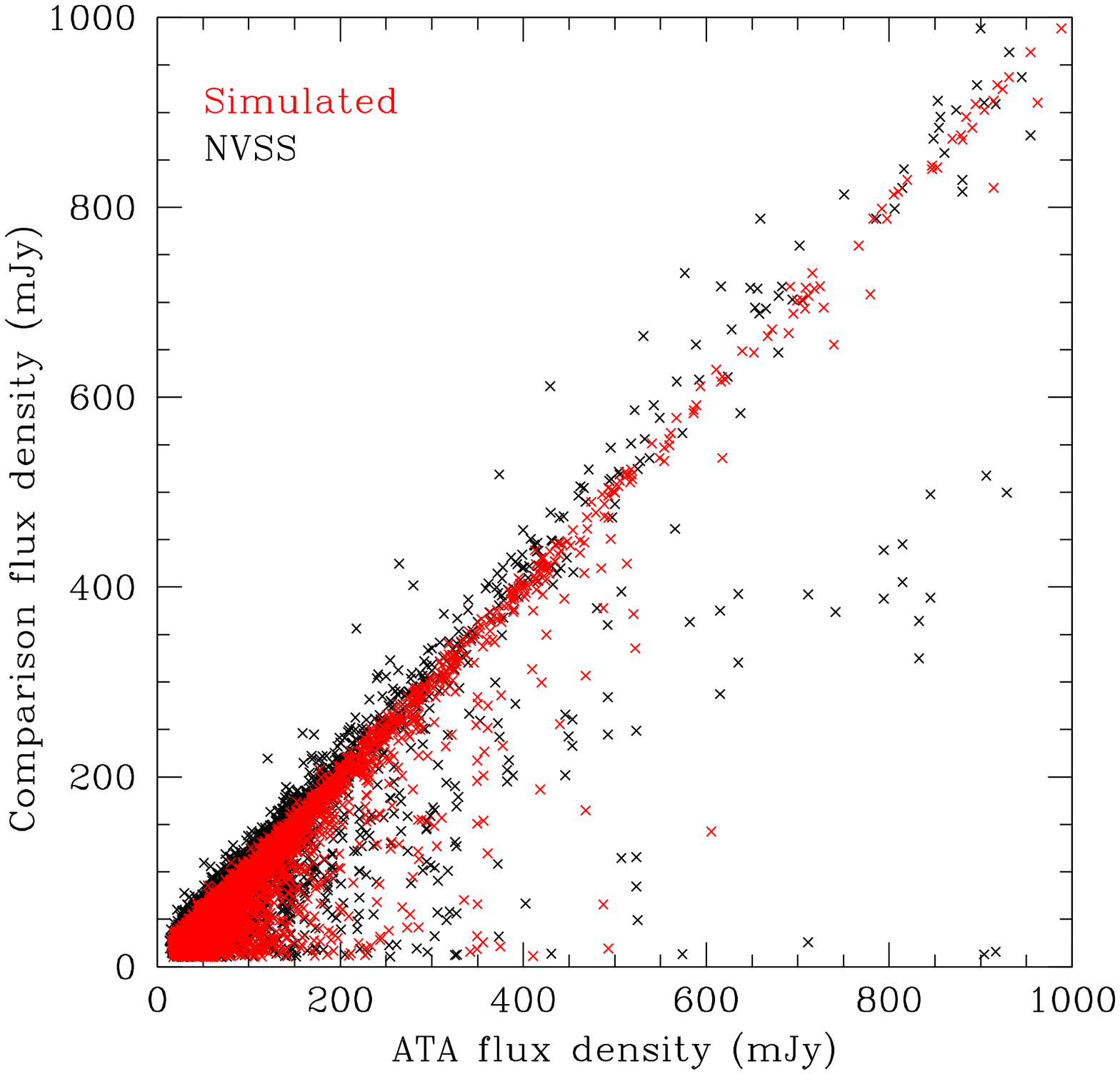}
\caption{\label{fig:realvfake}
The same data as in Fig.~\ref{fig:fluxmatch} (ATATS sources with NVSS matches, plotted in black) and Fig.~\ref{fig:fluxmatchfake} (retrieved flux densities for simulated sources inserted into ATATS, plotted in red), shown on a single plot to aid in comparison.
}
\end{figure}

As noted in \S~\ref{sec:realvfake}, simulated sources inserted into the image and then recovered in the catalog give a measure of the intrinsic scatter due to measurement error and confusion. When the ATATS catalog is compared to NVSS, additional scatter is partly due to intrinsic variability, which we attempt to quantify here. Fig.~\ref{fig:realvfake} shows the same data as in Figs.~\ref{fig:fluxmatch} and \ref{fig:fluxmatchfake} (for sources between 0 and 1\,Jy), this time on a single plot to aid comparison. Points inserted from the simulated catalog have the same resolution as ATATS, and do not cluster, unlike when comparing to NVSS, which has higher resolution and multi-component sources. As a result, the ATATS-NVSS match has some points (in the lower right quadrant of Fig.~\ref{fig:realvfake}) that are due to the ATA detecting a single object that is resolved by NVSS into several fainter components. Such sources are not present in the match to the simulated catalog. Even in the simulated catalog, however, an inserted source can still be matched erroneously to a nearby ATA source; these are the red points in Fig.~\ref{fig:realvfake} scattered well below the main cloud of points. So even when the resolutions are matched, and sources do not vary with time (the red points) there is still quite a lot of scatter just caused by measurement error and confusion. Clearly, though, there is some additional spread in the black points, which is at least in part due to variability of the sources.

We use the measured flux densities, \smi, and comparison flux densites, \sci\ to compute the Pearson product-moment correlation coefficient,

\begin{equation}
r = \frac{\sum_{i=1}^n (\smi - \sma) (\sci - \sca)}
{\sqrt{\sum_{i=1}^n (\smi - \sma)^2} \sqrt{\sum_{i=1}^n(\sci - \sca)^2}}
\end{equation}

for the sources shown in Figs.~\ref{fig:fluxmatch} (ATATS near NVSS positions), \ref{fig:fluxmatchatats} (NVSS near ATATS positions) and \ref{fig:fluxmatchfake} (simulated source retrieval); the values are $r = 0.863$, 0.969, and 0.973, respectively. When we smooth the NVSS catalog (as in Fig.~\ref{fig:fluxmatchsmooth}), the correlation is even better ($r = 0.985$), but we cannot apply this smoothing to the simulated catalog because there the sources have random positions. A clue to the additional scatter introduced by source variability is given by the comparison of $r$ for NVSS near ATATS and for retrieval of simulated sources, which shows that measurement uncertainties dominate.

\begin{figure}
\centering
\includegraphics[width=\linewidth,draft=false]{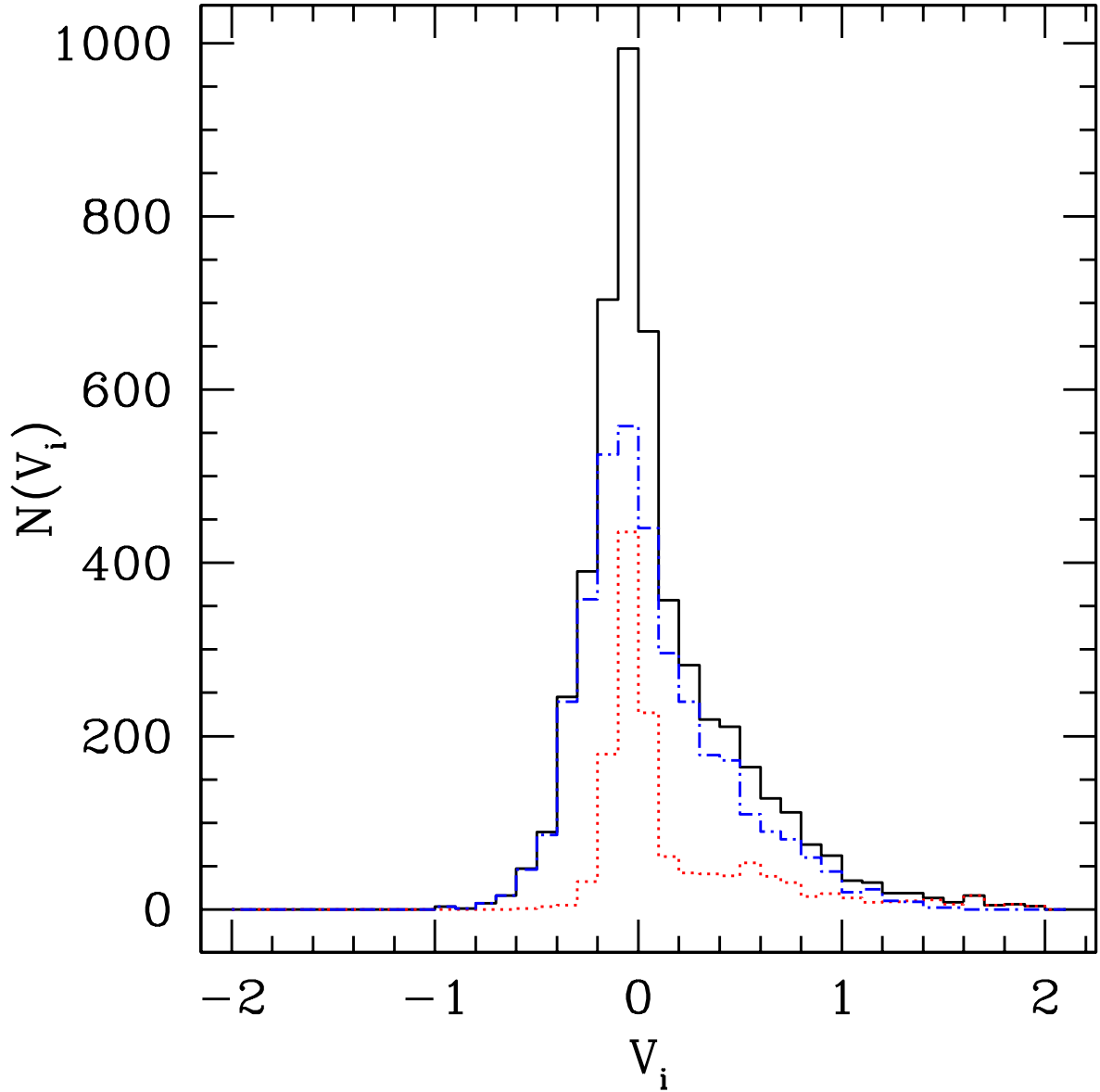}
\caption{\label{fig:vihist}
Histogram of the fractional flux variability, $V_i$ (see \S~\ref{sec:variability}) for the ATATS-NVSS match (black solid line). Also plotted are the histogram for ATATS sources fainter (blue dot-dashed line) and brighter (red dotted line) than 100\,mJy. Positive values of $V_i$ correspond to sources which are brighter in ATATS than in NVSS.
}
\end{figure}

We can examine this in more detail by computing the scatter in the points for the two samples. We define the fractional variability of each data point,

\begin{equation}
V_i = \frac{2 \times (\smi - \sci)}{\smi + \sci}
\end{equation}

and in Fig.~\ref{fig:vihist}, we plot a histogram in $V_i$ for the ATATS-NVSS match (for the sample as a whole, and split into subsamples of sources with $\satats \leq 100$\,mJy and $\satats > 100$\,mJy). The central peak of the histogram is narrower for the bright sources, showing that their fractional variability is lower. All the histograms show the asymmetry towards high $V_i$ due to the mismatch of bright ATATS and fainter NVSS sources discussed above. To obtain a measure of the true variability we wish to quantify the scatter in $V_i$ without being biased by these mismatched outliers, so we choose to measure the interquartile range of $V_i$, as opposed to the standard deviation. We compute the interquartile range, ${\rm IQR}_V$, in 26 flux density bins, with bin edges selected to place an equal number (188) of data points, \smi, in each bin.

\begin{figure}
\centering
\includegraphics[width=\linewidth,draft=false]{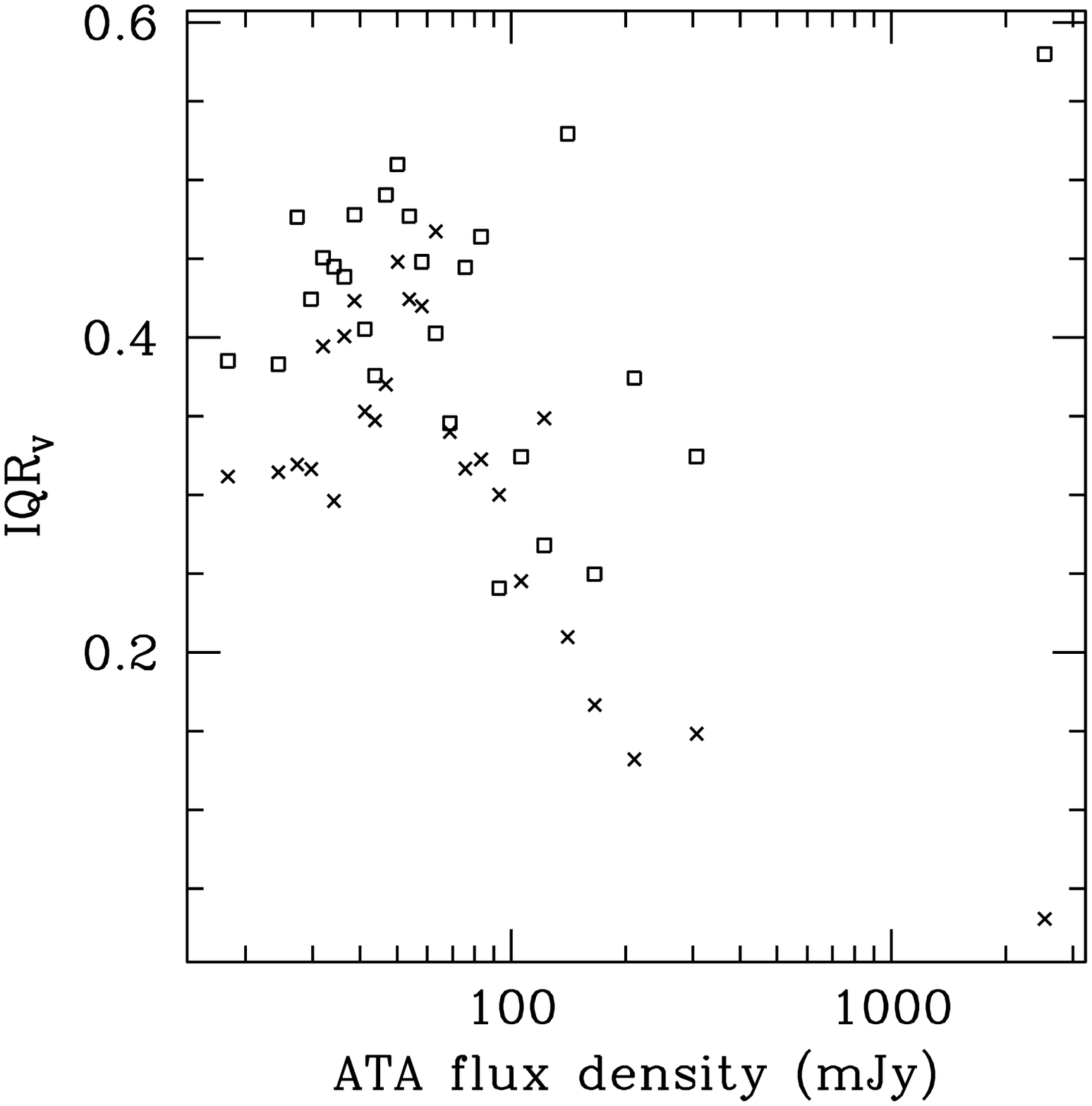}
\caption{\label{fig:sigmav}
${\rm IQR}_V$, the interquartile range of the fractional flux variability, $V_i$ (see \S~\ref{sec:variability}), computed for ATATS sources with NVSS matches (squares) and retrieved flux densities for simulated sources inserted into ATATS (crosses). Generally the scatter is larger for the real sources. 
}
\end{figure}

By examining the change in ${\rm IQR}_V$ as a function of flux density, we can quantify the spread of the points in the flux-flux plots. As well as computing ${\rm IQR}_V$ for the ATATS-NVSS match, we compute the same statistic comparing the measured flux densities, \smi, to the inserted simulated catalog values, \sci. We plot ${\rm IQR}_V$ for the two samples in Fig.~\ref{fig:sigmav}. Although there is quite a large amount of scatter in the points, we can see that generally the ATATS-NVSS match tends to show a little more variability than the simulated catalog, as discussed above, especially above $\sim 300$\,mJy, where the reliability of retrieval of simulated sources improves, but the scatter in the match to NVSS is higher. Some resolution effects undoubtedly remain, but at least some of this additional variability is presumably intrinsic. The fractional variability generally decreases towards higher flux densities (with the exception of the brightest bin, where the mismatched sensitivities and resolutions appear to dominate).

\begin{figure}
\centering
\includegraphics[width=0.85\linewidth,draft=false]{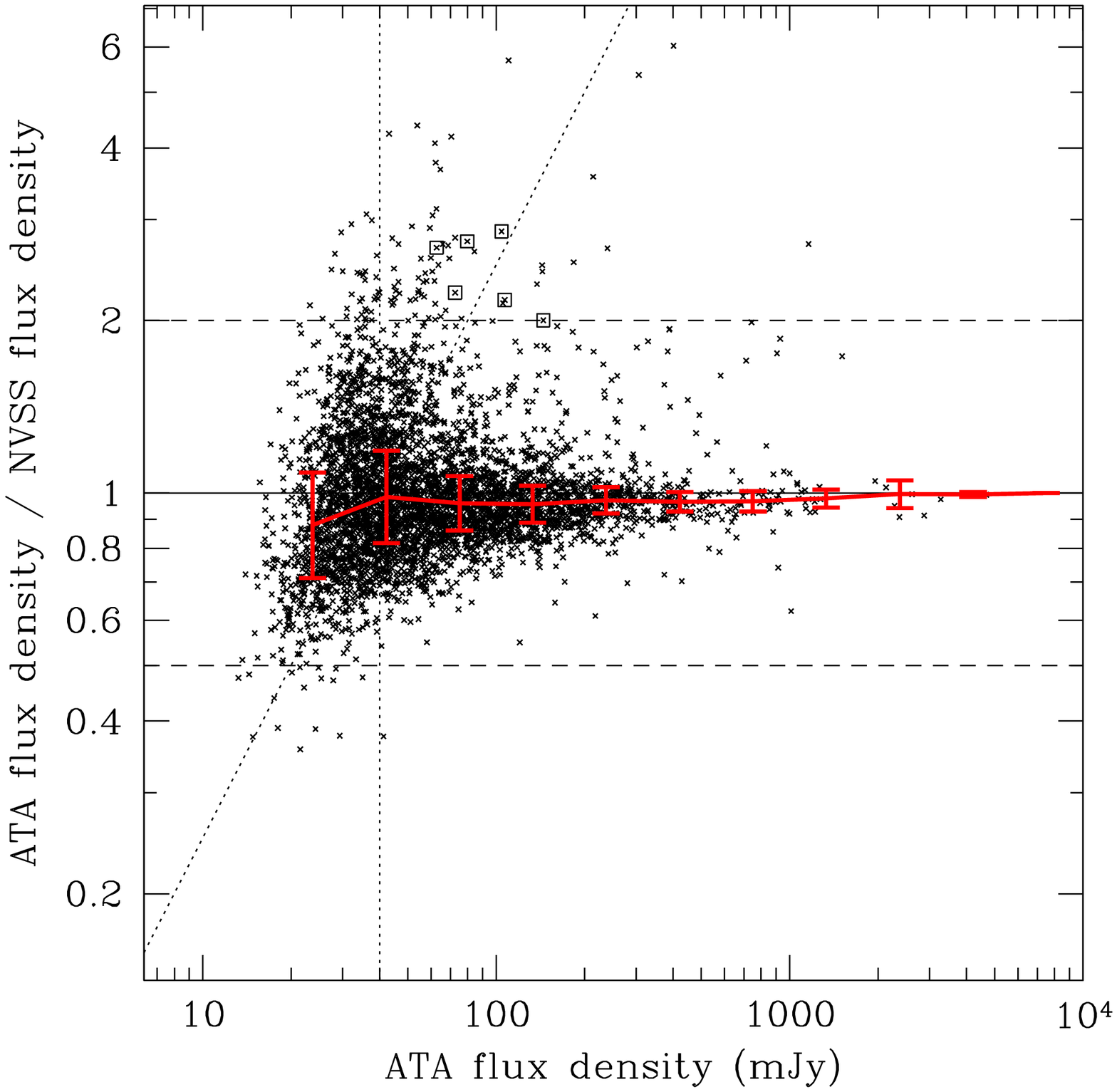}
\caption{\label{fig:factor2}
Comparison of the flux densities of sources from the ATATS catalog, with the sum of flux densities of all NVSS sources within 75\arcsec\ of the ATATS positions. The data are the same as those plotted in Fig.~\ref{fig:fluxmatchsmooth}, but here, in place of the NVSS flux density, we plot the ratio of the ATA to NVSS flux densities (black crosses). Sources whose flux density does not change from the smoothed NVSS measurement to the ATATS measurement lie on the horizontal solid 1:1 line. Sources whose fluxes change by a factor of 2 lie on the horizontal dashed 1:2 and 2:1 lines. The vertical dotted line shows the ATATS completeness limit, $\satats = 40$\,mJy. The slanted dotted line shows the same limit for NVSS, $\snvss = 40$\,mJy. The red solid line shows the median flux ratio in bins of 0.25 dex in \satats, and the associated error bars show the interquartile range of the flux density ratio.
We plot (as crosses enclosed by squares) the 6 sources above the ATATS completeness limit (vertical dotted line) which varied by a factor of 2 or more and which appeared relatively compact and isolated in both the ATATS and NVSS images (see \S~\ref{sec:factor2}).
}
\end{figure}

We take the ATATS and smoothed NVSS flux densities (as plotted in Fig.~\ref{fig:fluxmatchsmooth}), and plot their ratio as a function of ATATS flux density in Fig.~\ref{fig:factor2}. Again we see that the scatter around the 1:1 line increases towards fainter ATATS flux densities, suggestive of increased fractional variability for fainter sources. As discussed in \S~\ref{sec:symmetric}, an asymmetry in the point cloud can be seen, most obviously at flux densities fainter than the ATATS completeness limit (to the left of the dotted lines in Fig.~\ref{fig:factor2}. Above the completeness limit the point cloud is more symmetric about the 1:1 line, although there are still more outliers that appeared to get brighter from NVSS to ATATS than those that appeared to get fainter. 

We also plot in Fig.~\ref{fig:factor2} the median flux density ratio in bins of 0.25 dex in \satats. For sources brighter than the completeness limit, the median ratio of $\satats / \snvss$ is $0.97 \pm 0.10$\label{sec:bias}. This suggests that ATATS marginally underestimates the fluxes of NVSS sources. This may be due to a combination of the differing survey resolutions, the accuracy of the overall flux density calibration, and CLEAN bias \citep{nvss}. However, we do not expect CLEAN bias to be the dominant effect. The beam sidelobes are $< 10$\%\ of the main lobe (Fig.~\ref{fig:beamplot}), the RAPID software is designed not to CLEAN too deeply, and the combination of 12 snapshot observations into the master mosaic means that the \uv\ coverage is relatively good (Fig.~\ref{fig:uvcover}).

Some of the sources in Fig.~\ref{fig:realvfake} which exhibit more variability than the simulated (non-variable) sources are likely real variables, especially those which are significantly brighter in NVSS than in ATATS (because the resolution effects discussed above tend to increase scatter towards the lower right of the 1:1 line in this plot rather than towards the upper left). Some sources that appear highly variable turn out not to be on closer inspection, however, so we carefully examined the most extreme cases.

\begin{figure}
\centering
\includegraphics[width=0.3\linewidth,draft=false,bb=167 429 309 511]{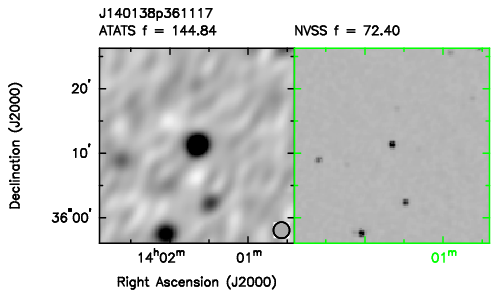}\hspace{0.03\linewidth}%
\includegraphics[width=0.3\linewidth,draft=false,bb=167 429 309 511]{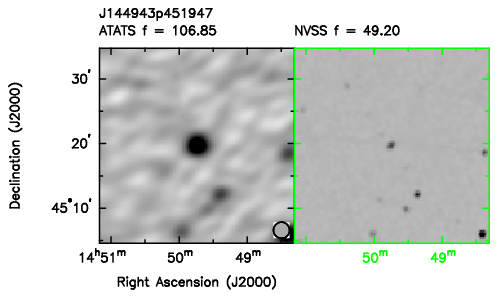}\hspace{0.03\linewidth}%
\includegraphics[width=0.3\linewidth,draft=false,bb=167 429 309 511]{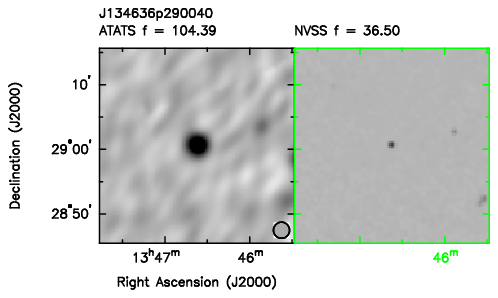}
\includegraphics[width=0.3\linewidth,draft=false,bb=167 429 309 511]{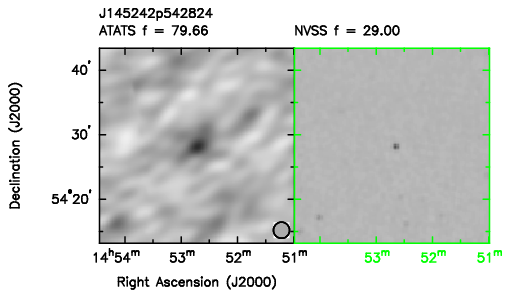}\hspace{0.03\linewidth}%
\includegraphics[width=0.3\linewidth,draft=false,bb=167 429 309 511]{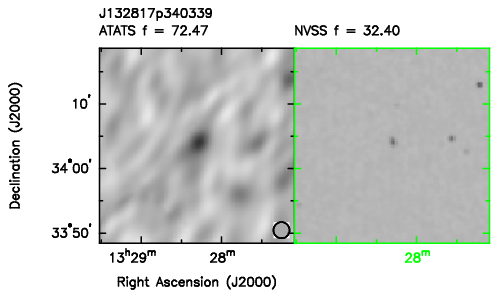}\hspace{0.03\linewidth}%
\includegraphics[width=0.3\linewidth,draft=false,bb=167 429 309 511]{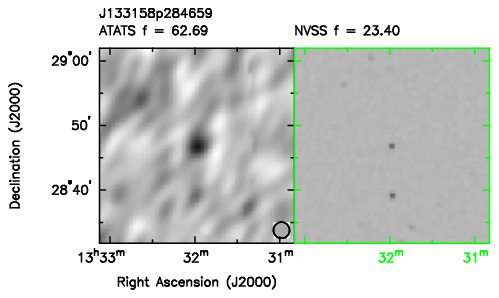}
\caption{\label{fig:varcand}
The six sources which we judged good candidates (\ie, those which appear compact and isolated, and are brighter than the ATATS completeness limit) to have varied by more than a factor 2 in flux density from NVSS to ATATS. \postsize\
}
\end{figure}

\begin{figure}
\centering
\includegraphics[width=0.3\linewidth,draft=false,bb=167 429 309 511]{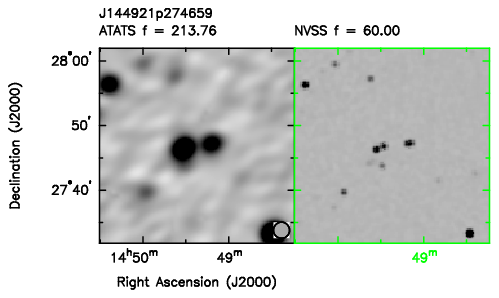}\hspace{0.03\linewidth}%
\includegraphics[width=0.3\linewidth,draft=false,bb=167 429 309 511]{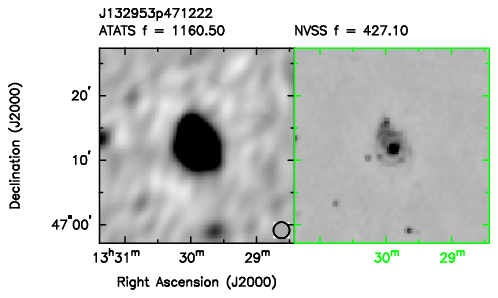}\hspace{0.03\linewidth}%
\includegraphics[width=0.3\linewidth,draft=false,bb=167 429 309 511]{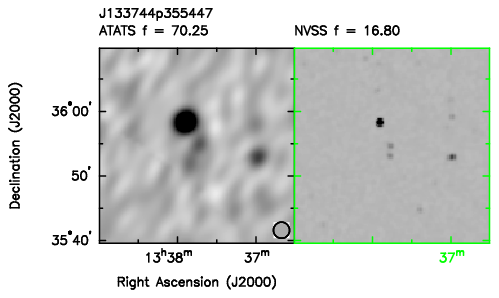}
\caption{\label{fig:novarcand}
Examples of some of the 61 candidates that appeared to have varied by more than a factor 2 in flux density from NVSS to ATATS, but have complex morphologies. Such sources are prone to resolution effects and less likely to be truly highly variable than are isolated sources. \postsize\
}
\end{figure}

We examined postage stamps from the ATATS mosaic and from NVSS for the 67 sources brighter than the ATATS completeness limit which have measured fluxes a factor of 2 or more different (either brighter or fainter) from their flux densities as measured from the smoothed NVSS catalog. We judged that only 6 of the sources were relatively compact and isolated in both NVSS and ATATS, and may therefore have varied by a factor of 2 or more in the 15 years between epochs. These sources are marked by squares in Fig.~\ref{fig:factor2}. Postage stamps for these sources are shown in Fig.~\ref{fig:varcand}. The remaining 61 candidates had complex morphologies or near neighbors (Fig.~\ref{fig:novarcand}). These are most likely cases where multiple NVSS sources were not deblended successfully by the lower-resolution ATATS (see \S~\ref{sec:nobrighttransients}), although we cannot rule out that some may really be highly variable -- however, as stated in \S~\ref{sec:symmetric}, we would expect such variations to be symmetrically distributed about the 1:1 line if real.

We plot the candidate highly variable sources on the plot of flux ratio versus flux density in Fig.~\ref{fig:factor2}.\label{sec:factor2} As can be seen, no highly variable sources are seen to get {\em fainter} from NVSS to ATATS, suggesting that most apparently highly variable sources appear so because of resolution effects. Of the 6 good candidates, 2 were brighter than $\snvss = 40$\,mJy (\ie, to the right of the slanted dotted line in Fig.~\ref{fig:factor2}). A search of the NASA Extragalactic Database shows that the five brightest sources of the six are known QSOs or radio galaxies (Table~\ref{tab:varcand}). We judge that the faintest is also likely a QSO (although not previously classified as such) due to its compact radio morphology and variable radio flux.

\begin{deluxetable}{rrrrrrl}
\rotate
\tablewidth{0pt}
\tabletypesize{\scriptsize}
\tablecaption{\label{tab:varcand} Sources which are likely to have varied by a factor 2 between NVSS and ATATS}
\tablehead {
\colhead{Catalog ID} &
\colhead{RA (J2000)} &
\colhead{Dec (J2000)} &
\colhead{\satats\ (mJy)} &
\colhead{$\sigma_{\rm bg}$(mJy)} &
\colhead{\snvss\ (mJy) \tablenotemark{a}} &
\colhead{Probable identification} 
}
\startdata
ATATS\,J140138+361117&	\hms{14}{01}{38.8} & \dms{36}{11}{17} & 144.8 & 5.2 & 72.4 &MG2\,J140141+3611 ($z=0.507$ BL Lac)\\
ATATS\,J144943+451947&	\hms{14}{49}{43.8} & \dms{45}{19}{47} & 106.8 & 4.4 & 49.2 &87GB\,144756.5+453159 ($z=1.635$ QSO)\\
ATATS\,J134636+290040&	\hms{13}{46}{36.9} & \dms{29}{00}{40} & 104.4 & 3.2 & 36.5 &MG2\,J134636+2900 ($z=2.721$ QSO)\\
ATATS\,J145242+542824&	\hms{14}{52}{42.6} & \dms{54}{28}{24} & 79.7 & 4.7 & 29.0 &SDSS\,J145239.17+542808.6 ($z=2.838$ QSO) \\
ATATS\,J132817+340339&	\hms{13}{28}{18.0} & \dms{34}{03}{39} & 72.5 & 4.9 & 32.4 &7C\,1326+3419\\
ATATS\,J133158+284659&	\hms{13}{31}{58.1} & \dms{28}{46}{59} & 62.7 & 6.0 & 23.4 & FIRST\,J133158.7+28465\\
\enddata
\tablenotetext{a}{Sum of integrated flux densities of NVSS sources within 75\arcsec of ATATS position}
\tablecomments{The six sources which we judged good candidates (\ie, those which appear compact and isolated, and are brighter than the ATATS completeness limit) to have varied by more than a factor 2 in flux density from NVSS to ATATS. Associations (with redshifts where available) are from the NASA Extragalactic Database.}
\end{deluxetable}

\subsection{Transient Statistics}

By combining the data from all 12 epochs to make the master mosaic, we should gain a factor 3.5 in RMS sensitivity (which scales as the square root of the observing time), but in practice we gain somewhat more in image fidelity due to the poor \uv\ coverage of the individual snapshots. However, this comes at the expense of sensitivity to short duration transients. A transient with a characteristic timescale much shorter than the several days between ATATS epochs, and hence which appears only in a single epoch, will have a flux density in the master mosaic a factor $\sim 12$ less than in the image made from the epoch in which it occurred. So we expect that by comparing catalogs generated from the 12 single-epoch images, we will detect any such transients with $\sim 3.5$ times better signal-to-noise than we are able to by comparing the master mosaic and NVSS. The analysis of the constraints on transients and variability from epoch to epoch will be the subject of a forthcoming paper \citep{paperii}. For transients with characteristic timescales longer than the 81 days between epochs 1 and 12, but shorter than the $\sim 15$\,yr between NVSS and ATATS, the master mosaic catalog compared to NVSS (as discussed in this paper) provides the best sensitivity.

\begin{figure}
\centering
\includegraphics[width=0.3\linewidth,draft=false,bb=167 429 309 511]{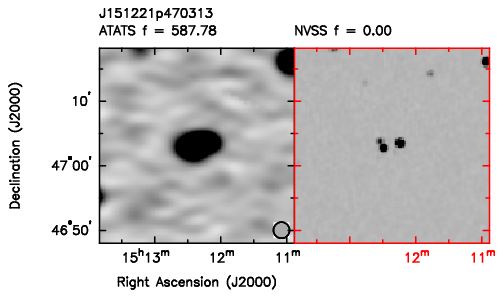}\hspace{0.03\linewidth}%
\includegraphics[width=0.3\linewidth,draft=false,bb=167 429 309 511]{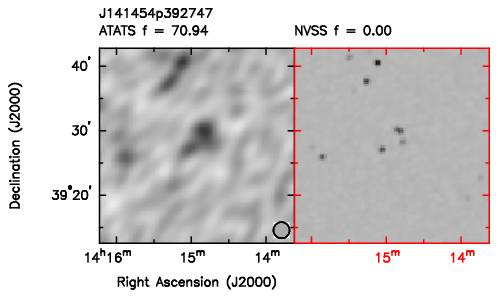}\hspace{0.03\linewidth}%
\includegraphics[width=0.3\linewidth,draft=false,bb=167 429 309 511]{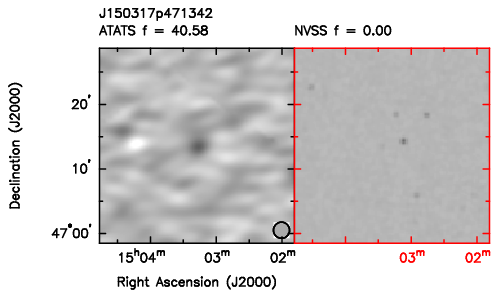}
\caption{\label{fig:transcand}
Examples of some of the 39 transient candidates with $\satats \geq 40$\,mJy and $\snvss = 0$. None are truly transients (\S~\ref{sec:nobrighttransients}). \postsizetwo\
}
\end{figure}

To look for such transients, we matched the final ATATS catalog of 4408 sources with the culled NVSS catalog (\S~\ref{sec:cull}) of 10729 sources, once again with a 75\arcsec\ match radius. 4333 of the ATATS sources had a match in NVSS. Of the remaining 75 sources without an NVSS match, 39 are above 40\,mJy, the ATATS 90\%\ completeness limit. We examined postage stamp images for these 39 sources (from the ATATS mosaic and from NVSS) and concluded that none were real transient sources.\label{sec:nobrighttransients} In the majority of cases, these were sources which consisted of multiple NVSS sources (usually as part of a double source -- presumably a radio galaxy) which were not successfully deblended by ATATS. Examples of some of these sources are shown in Fig.~\ref{fig:transcand}.

NVSS is complete to a much fainter flux density limit than ATATS, so we can be more confident that an ATATS source without an NVSS counterpart is a transient than we could be confident that an NVSS source of similar brightness without an ATATS counterpart is a transient. Nevertheless, we searched for NVSS sources within the ATATS field without an ATATS counterpart. We limited ourselves to the 978 NVSS sources brighter than $\snvss = 120$\,mJy (\ie, three times the measured ATATS 90\%\ completeness limit) where ATATS is over 97\%\ complete (\S~\ref{sec:nomatch}). We examined postage stamp images of the 26 NVSS sources brighter than this limit with no ATA counterpart. All are sources with complex morphologies, and in each case it is clear from the images that the NVSS source does in fact have a counterpart in ATATS. Although comparing the catalogs using a 75\arcsec\ match radius results in only 97\%\ completeness, the fact that fewer than 1 out of 978 NVSS sources brighter than 120\,mJy are truly missed by ATATS when the morphologies of complex sources are examined means that we can be $> 99.9$\%\ confident that no real 120\,mJy sources were undetected in the ATATS images.\label{sec:99.9complete}

We can use the lack of transients seen when comparing ATATS to NVSS to place constraints on the population of transient sources brighter than $\sim 40$\,mJy. \label{sec:transrates} We see no sources in ATATS without a counterpart in NVSS (after visual inspection of the 39 candidates without a match within 75\arcsec). Poisson statistics give the probability

\begin{equation}
P(n) = \frac{e^{-r}r^n}{n!}
\end{equation}

of detecting $n$ transients given that $r$ transients are expected over the survey area. Poisson $2\sigma$ confidence intervals correspond to $P(n) = 1 - 0.95 = 0.05$, and given that we detect $n = 0$ sources in the field, we derive a $2\sigma$ upper limit on the transient rate of $r = -{\rm ln}\,0.05 =  3$ sources in the 690~\sqdeg\ survey, or 0.004\,deg$^{-2}$.

\begin{figure}
\centering
\includegraphics[width=\linewidth,draft=false]{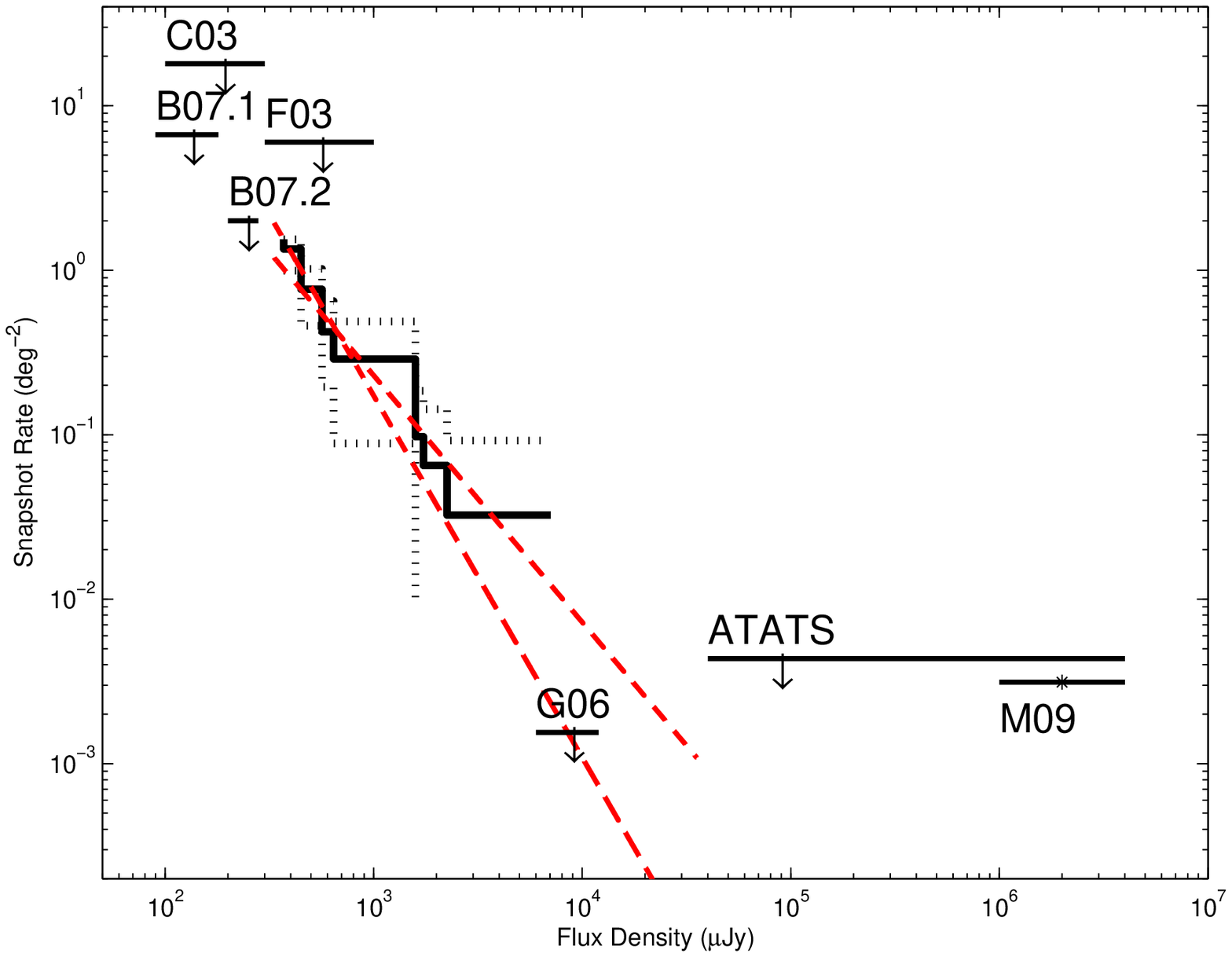}
\caption{\label{fig:rate}
Cumulative two-epoch source density for radio transients as a
function of flux density, based on Fig.~9 of \citet{bower}. The solid black line shows the rate measured for transients with characteristic timescales $< 7$~days from \citet{bower}, while the broken black lines show the $2 \sigma$ upper and lower bounds. The red dashed lines show $S^{-1.5}$ and
$S^{-2.2}$ curves. The arrows show $2 \sigma$ upper limits for
transients from \citeauthor{bower} with a 1\,yr timescale (labeled B07.1) and a 2 month timescale (labeled B07.2), and for
transients from the comparison of the 1.4\,GHz NVSS and FIRST surveys by \citet{galyam}, labeled G06; from the \citet{carilli:03} survey (labeled C03); from the \citet{frail:03} survey (labeled F03); and from the \citet{matsumura:09} survey (labeled M09). The upper limits from ATATS (\S~\ref{sec:transrates}) are only marginally consistent with the \citeauthor{matsumura:09} detections. Constraints from data from individual ATATS epochs will be discussed in \citet{paperii}.
}
\end{figure}

Fig.~\ref{fig:rate} shows the constraints on transient rates from some of the surveys discussed in \S~\ref{sec:archival}, compared to the constraint from the comparison of ATATS and NVSS. It is hard to reconcile the steeply falling source counts suggested by the \citet{galyam} and ATATS results with the detections of 9 transients brighter than 1\,Jy reported by \citet{matsumura:09} unless these transients make up a different population with a very steep cutoff below 1\,Jy. The ATATS $2 \sigma$ upper limit is marginally consistent with the M09 results above 1\,Jy, but if the \citeauthor{matsumura:09} transients have flat or rising source counts towards fluxes $< 1$\,Jy we would expect to see them in ATATS. However, it should be noted that the characteristic timescale of the transients reported by \citeauthor{matsumura:09} is minutes to days, whereas our averaging of 12 ATATS epochs means that transients with durations shorter than a few days (\ie, those which might be present in a single ATATS epoch but not in the other 11) will have mean flux densities reduced by a factor 12 (and signal-to-noise by a factor 3.5) relative to the single-epoch values. Nevertheless, if there are transients which reach average flux densities brighter than 12 times our completeness limit (\ie, $\sim 500$\,mJy and above) for at least one minute (the integration time for each individual ATATS epoch), they should appear as sources at least as bright as the completeness limit (\ie, averaged down by up to a factor 12 if undetected at other epochs) in the master mosaic. No such sources are seen.

We will examine transient statistics from the images and catalogs from individual epochs in \citet{paperii}.

\section{Summary} 

Analysis of the combined 12-epoch ATATS dataset shows that the ATA is able to produce accurate images of the radio sky, with positions and fluxes matching the legacy NVSS survey at the same frequency. A search for sources seen in ATATS but not present in NVSS (within a 75\arcsec\ match radius) turned up 39 candidates above the 40\,mJy ATATS completeness limit, but upon inspection, none are plausible transient sources -- the majority are double sources in NVSS whose components are more than 75\arcsec\ from the ATATS position. This enables us to place limits on the transient population of fewer than 0.004 transients deg$^{-2}$, for transients brighter than 40\,mJy with characteristic timescales $\gtrsim 3$~months. The same limit applies to transients brighter than $\sim 500$\,mJy with characteristic timescales of minutes or longer. Such sources will be discussed further in the next ATATS paper.

\acknowledgments

The authors would like to acknowledge the generous support of the Paul
 G. Allen Family
 Foundation, which has provided major support for the design, construction,
 and operation of
 the ATA. Contributions from Nathan Myhrvold, Xilinx Corporation, Sun
 Microsystems,
 and other private donors have been instrumental in supporting the ATA.
 The ATA has been
 supported by contributions from the US Naval Observatory in addition
 to National Science
 Foundation grants AST-050690 and AST-0838268.
 
 We thank the anonymous referee for a careful and thorough reading of our paper, and for constructive suggestions which improved the manuscript.


\begin{thebibliography}{}

\bibitem[Aller et al.(1992)]{aller:92} Aller, M.~F., Aller, H.~D., \& Hughes, P.~A.\ 1992, \apj, 399, 16 
\bibitem[Archibald et al.(2009)]{archibald} Archibald, A.~M., et al.\ 2009, Science, 324, 1411 
\bibitem[Arshakian et al.(2010)]{arshakian:10} Arshakian, T.~G.,  Le{\'o}n-Tavares, J., Lobanov, A.~P., Chavushyan, V.~H., Shapovalova, A.~I., Burenkov, A.~N., \& Zensus, J.~A.\ 2010, \mnras, 401, 1231 
\bibitem[Baars et al.(1977)]{baars} Baars, J.~W.~M., Genzel, R., Pauliny-Toth, I.~I.~K., \& Witzel, A.\ 1977, \aap, 61, 99 
\bibitem[Berger et al.(2001)]{berger:01} Berger, E., et al.\ 2001, \nat, 410, 338
\bibitem[Bower et al.(2007)]{bower} Bower, G.~C., Saul, D., Bloom, J.~S., Bolatto, A., Filippenko, A.~V., Foley, R.~J., \& Perley, D.\ 2007, \apj, 666, 346 
\bibitem[Brunthaler et al.(2009)]{m82sn} Brunthaler, A., Menten, K.~M., Reid, M.~J., Henkel, C., Bower, G.~C., \& Falcke, H.\ 2009, \aap, 499, L17
\bibitem[Burke-Spolaor et al.(2010)]{bs}Burke-Spolaor, S., Bailes, M., Ekers, R.~D., Macquart, J.-P., \& Crawford, F.\ 2010, in prep. 
\bibitem[Carilli et al.(2003)]{carilli:03} Carilli, C.~L., Ivison, R.~J., \& Frail, D.~A.\ 2003, \apj, 590, 192 
\bibitem[Cognard et al.(1996)]{cognard:96} Cognard, I., Shrauner, J.~A., Taylor, J.~H., \& Thorsett, S.~E.\ 1996, \apjl, 457, L81
\bibitem[Condon et al.(1998)]{nvss} Condon, J.~J., Cotton, W.~D., Greisen, E.~W., Yin, Q.~F., Perley, R.~A., Taylor, G.~B., \& Broderick, J.~J.\ 1998, \aj, 115, 1693 
\bibitem[Croft et al.(2010b)]{paperii} Croft, S., et al.\ 2010b, in prep.
\bibitem[Fiedler et al.(1987)]{ese} Fiedler, R.~L., Dennison, B., Johnston, K.~J., \& Hewish, A.\ 1987, \nat, 326, 675 
\bibitem[Frail et al.(2000)]{frail:00} Frail, D.~A., Waxman, E., \& Kulkarni, S.~R.\ 2000, \apj, 537, 191 
\bibitem[Frail et al.(2003)]{frail:03} Frail, D.~A., Kulkarni, S.~R., Berger, E., \& Wieringa, M.~H.\ 2003, \aj, 125, 2299
\bibitem[Gal-Yam et al.(2006)]{galyam} Gal-Yam, A., et al.\ 2006, \apj, 639, 331 
\bibitem[G{\"u}del(2002)]{gudel:02} G{\"u}del, M.\ 2002, \araa, 40, 217 
\bibitem[Harp et al.(2010)]{harp} Harp, G., et al.\ 2010, IEEE Proceedings, submitted. 
\bibitem[Hallinan et al.(2007)]{hallinan:07} Hallinan, G., et al.\ 2007, \apjl, 663, L25 
\bibitem[Hopkins et al.(2005)]{hopkins} Hopkins, P.~F., et al.\ 2005, \apjl, 625, L71  
\bibitem[Hull et al.(2010)]{hull} Hull, C., et al.\ 2010, in prep.
\bibitem[Falcke et al.(1999)]{falcke:99} Falcke, H., et al.\ 1999, \apjl, 514, L17 
\bibitem[Jackson et al.(1989)]{jackson:89} Jackson, P.~D., Kundu, M.~R., \& White, S.~M.\ 1989, \aap, 210, 284 
\bibitem[Jannuzi \& Dey(1999)]{ndwfs} Jannuzi, B.~T., \& Dey, A.\ 1999, ASP Conf.~Ser.~193: The Hy-Redshift Universe: Galaxy Formation and Evolution at High Redshift, 193, 258 
\bibitem[Keating et al.(2009)]{keating:aas} Keating, G., Barott, J., \& Allen Telescope Array Team 2009, American Astronomical Society Meeting Abstracts, 214, 601.06 
\bibitem[Koopmans et al.(2003)]{microlensing} Koopmans, L.~V.~E., et al.\ 2003, \apj, 595, 712
\bibitem[Koz{\l}owski et al.(2010)]{sdwfsvar} Koz{\l}owski, S., et al.\ 2010, \apj, 716, 530  
\bibitem[Lazio et al.(2008)]{lazio} Lazio, T.~J.~W., et al.\ 2008, \apjl, 672, L115 
\bibitem[Lorimer et al.(2007)]{lorimer:07} Lorimer, D.~R., Bailes, M., McLaughlin, M.~A., Narkevic, D.~J., \& Crawford, F.\ 2007, Science, 318, 777 
\bibitem[MacMahon \& Wright(2009)]{hex7} MacMahon, D., \& Wright, M. (2009), ``Antenna Gain, Pointing, Primary Beam and Bandpass Calibration'', ATA Memo Series 83
\bibitem[Marscher et al.(2002)]{marscher:02} Marscher, A.~P., Jorstad, S.~G., G{\'o}mez, J.-L., Aller, M.~F., Ter{\"a}sranta, H., Lister, M.~L., \& Stirling, A.~M.\ 2002, \nat, 417, 625 
\bibitem[Matsumura et al.(2009)]{matsumura:09} Matsumura, N., et al.\ 2009, \aj, 138, 787 
\bibitem[McLaughlin et al.(2006)]{rrats} McLaughlin, M.~A., et al.\ 2006, \nat, 439, 817
\bibitem[Miller-Jones et al.(2009)]{mj:09} Miller-Jones, J.~C.~A. et al.\ 2009, \mnras, 394, 309
\bibitem[Ofek et al.(2010)]{ofek:10} Ofek, E.~O., Breslauer, B., Gal-Yam, A., Frail, D.,Kasliwal, M.~M., Kulkarni, S.~R., \& Waxman, E.\ 2010, \apj, 711, 517 
\bibitem[Sagar et al.(2004)]{sagar} Sagar, R., et al.\ 2004, \mnras, 348, 176
\bibitem[Saikia et al.(2006)]{saikia} Saikia, D.~J., et al.\ 2006, \mnras, 366, 1391 
\bibitem[Sargent \& Welch(1993)]{sargent} Sargent, A.~I., \& Welch, W.~J.\ 1993, \araa, 31, 297 
\bibitem[Sault et al.(1995)]{miriad} Sault, R.~J., Teuben, P.~J., \& Wright, M.~C.~H.\ 1995, Astronomical Data Analysis Software and Systems IV, 77, 433 
\bibitem[Seymour et al.(2008)]{seymour:08} Seymour, N., et al.\ 2008, \mnras, 386, 1695
\bibitem[Spergel et al.(2003)]{wmap} Spergel, D.~N., et al.\ 2003, \apjs, 148, 175
\bibitem[Welch et al.(2009)]{welch} Welch, J., et al.\ 2009, IEEE Proceedings, 97, 1438 
\bibitem[Williams \& Bower(2010)]{williams} Williams, P.~K.~G., \& Bower, G.~C.\ 2010, \apj, 710, 1462 

\end{thebibliography}
\end{document}